\documentclass[11pt, a4paper]{article}
\usepackage{amsmath}
\usepackage{amssymb}
\usepackage{geometry}
\usepackage{graphicx}
\usepackage[hidelinks]{hyperref}
\usepackage{url}
\usepackage{color}
\usepackage{bm}
\usepackage{cite}
\usepackage{authblk}
\usepackage[utf8]{inputenc}
\usepackage{booktabs} % For professional-quality tables

\DeclareMathOperator{\sgn}{sgn}
\hypersetup{
	colorlinks=true,
	linkcolor=blue,
	filecolor=magenta,      
	urlcolor=cyan,
	citecolor=red
}

\newcommand{\red}{\textcolor{black}}
\title{Harmonic Theory of Behavior}

\author[1,2,3]{Mohammad Salahshour\thanks{msalahshour@ab.mpg.de}}
\author[1,2,3]{Iain D. Couzin}

\affil[1]{Department of Collective Behaviour, Max Planck Institute of Animal Behavior, Konstanz, Germany}
\affil[2]{Centre for the Advanced Study of Collective Behaviour, University of Konstanz, Konstanz, Germany}
\affil[3]{Department of Biology, University of Konstanz, Konstanz, Germany}

\date{April 24, 2026}
\begin{document}
	
	\maketitle

	\begin{abstract}
		Traditional models of collective behavior rely on prescribed interaction rules, leaving unresolved the question of how behavior arises from neural representations of space. Here, we develop a first-principles theory in which movement, decision-making, and collective organization emerge by coarse-graining fast neural dynamics on a topological representation of directional space. For a ring manifold encoding heading, this reduction yields a macroscopic theory of behavior: a decision landscape over directions that admits a harmonic decomposition. In this framework, behavior is governed by a spectral organization of directional information rather than ad hoc rules. We demonstrate that target-seeking, avoidance, choice, spatial decision-making, and diverse forms of collective motion arise as distinct organizations of the underlying harmonic landscape. The theory unifies neural representations, behavioral decisions, and collective dynamics in a single mathematical description.
		
	\end{abstract}

	\section{Introduction}

For over half a century, collective behavior in biological systems---from insect swarms to fish schools and bird flocks---has been understood through an order paradigm rooted in statistical physics \cite{Sumpter2006,Giardina2008,Vicsek2012}. Often summarized by the maxim ``more is different" \cite{Anderson1972}, this perspective holds that when many interacting units are coupled together, new macroscopic laws emerge that cannot be trivially inferred from the properties of individual units alone. In biological collectives, this perspective has been formalized in prominent models of collective motion \cite{Vicsek1995,Toner1995,Couzin2005,Ballerini2008,Ginelli2010}, showing how collective motion can arise from simple local interactions \cite{Couzin2007,Vicsek2012}. Within this framework, organisms are typically modeled as self-propelled particles that obey heuristic rules such as aligning with neighbors or moving toward a local center of mass \cite{Ramaswamy2010,Vicsek2012,Marchetti2013,Reynolds1987,Couzin2002}.
	
These rule-based approaches have successfully demonstrated that global coordination can be achieved without centralized control. Yet they leave a deeper biological question open: what mechanisms generate the apparent rules themselves? Rules of interaction describe behavioral outcomes, but not how nervous systems transform perception into action \cite{Reynolds1987,Couzin2003,Ouellette2021,Gautrais2012,Tinbergen1963,Krakauer2017}. In such models, the heading vector is treated as the fundamental dynamical variable of the organism, rather than as the result of an internal sensory and neural computation \cite{Tinbergen1963,Krakauer2017,Ouellette2021}. As a consequence, the organism’s decision architecture—and more generally the link between neural representations of space and behavior—remains outside the explanatory framework \cite{GomezMarin2014,Krakauer2017,Ouellette2021}. This has created a gap between the neuroscience of perception and the ecology of behavior, while also introducing a potential circularity in which organizational principles of behavior are assumed rather than explained \cite{Ouellette2021,Salahshour2025,Sayin2025}.
	
A central missing ingredient in such accounts is the internal representation of space. Animals do not make directional decisions by reacting to neighbors or targets as isolated objects; they first encode spatial information within structured neural manifolds. These internal representations provide a topological mapping of space that preserves neighborhood structure, so that nearby locations or directions in the environment remain nearby in neural state space. Empirical work across taxa has revealed several such topological mappings of space, including grid-like \cite{Moser2008,Hafting2005}, toroidal \cite{Gardner2022}, and ring-like manifolds \cite{Kim2017,Taube2007}. Although these architectures differ in geometry, they share the same core principle: environmental information is represented on neural manifolds that can preserve local adjacencies between the represented and neural spaces, and behavior arises from computations performed on such manifolds (Fig. \ref{Fig0}\textbf{A}). This observation suggests a broader reformulation of the problem of behavior. Rather than asking how collective order emerges from rules of interaction, we ask how macroscopic behavior emerges from environmental information processing on such a neural manifold.
	
In this paper, we focus on a ring manifold ($S^1$) representing directional space, as an evolutionarily preserved, biologically widespread \cite{Kim2017,Seelig2015,Sarel2017,Finkelstein2015}, and a mathematically tractable map of directional information in two-dimensional space \cite{Zhang1996,Seelig2015,Finkelstein2015,Taube2007,Pfeiffer2014,Kim2017,Sarel2017,Mussells2024,Westeinde2024,Wilson2023}. This allows us to reformulate the problem of behavior: rather than starting from prescribed interaction rules, we ask how macroscopic behavior emerges from environmental information processing on a topological mapping of directional space. We demonstrate that this perspective can bring diverse aspects of behavior, from individual spatial decision-making to collective behavior, under a single, unified framework.
	
By doing so, we introduce the Harmonic Theory of Behavior, a first-principles neuro-ecological framework that bridges the neuroscience of perception and the ecology of behavior \cite{Tinbergen1963,GomezMarin2014,Krakauer2017,Ouellette2021}. The theory begins from a simple premise: living organisms are perceptual agents equipped with neural circuits that encode environmental structure \cite{Moser2008,Hafting2005,Gardner2022,Finkelstein2015,Kim2017,Taube2007}, and act on the basis of those internal representations \cite{Tinbergen1963,Krakauer2017,Salahshour2025}. Macroscopic behavior, therefore, arises not simply from interactions among individuals---as traditionally framed \cite{Anderson1972,Ramaswamy2010,Vicsek2012,Marchetti2013}---but from how each individual represents and evaluates environmental information when selecting actions. In this view, individual and collective behavior emerge from the transformation of environmental information within these structured representations and from the coupling of these transformations across organisms through interaction.
	
At the macroscopic level, this theory is analogous to a thermodynamic description of behavior. Just as thermodynamics describes large-scale properties of matter by integrating out microscopic molecular dynamics \cite{Pathria2017,Kardar2007,Zwanzig2001}, the Harmonic Theory derives macroscopic equations of motion for individuals and collectives by integrating out the fast neural dynamics of underlying decision circuits \cite{Zwanzig1960,Zwanzig2001,Gardiner2004}. By treating the neural state as the microscopic substrate and the behavioral trajectory as the macroscopic observable \cite{Zwanzig2001,Kardar2007}, we establish a statistical mechanical link between the two scales. 

This reduction reveals that the ``rules'' of collective behavior can emerge via an effective Hamiltonian, a mathematical function that naturally assigns a relative preference to each possible movement direction. In biological terms, this function describes how the nervous system evaluates alternative headings at any given moment. We refer to this structured mapping from direction to preference as a decision landscape. Importantly, this landscape can be decomposed into a set of structured components, or harmonics, that together shape the decision landscape. These harmonics do not correspond directly to behavioral rules. Rather, they describe how directional information is distributed across possible headings. The combined harmonic structure determines the curvature and stability of the decision landscape. 
	
At the individual level, this harmonic computation of information gives rise to preferred directions, symmetry-breaking, bifurcations, and spatial pattern formation. At the collective level, it identifies fundamental drivers of consensus and conflict based on the parity of harmonic modes and the geometry of space. The modulation of these harmonics entails diverse forms of collective behavior, such as synchronization, rotational milling, fission-fusion dynamics, and ordered motion. This unified framework reveals that ``rules'' of individual and collective behavior, rather than being imposed as primitive interaction rules, can emerge from information processing on a topological representation of space.

\section{Results}
	
\subsection{The Harmonic Theory of Behavior}
	
We propose a general framework for behavior based on a simple idea: organisms do not respond to the world using fixed behavioral rules, but by first representing sensory information internally and then acting on that representation \cite{Krakauer2017,GomezMarin2014,Wilson2023}. In our framework, movement emerges from how environmental information is encoded, filtered, and transformed into action by the nervous system \cite{Wilson2023,Mussells2024,Westeinde2024}. This allows behavior to be described at a macroscopic level, while still remaining connected to the underlying biological mechanism.
	
The theory can be derived in two complementary ways: \emph{axiomatically}, by starting from general assumptions about how organisms encode and process spatial information, and \emph{mechanistically}, by starting from an explicit microscopic neural model and coarse-graining its fast dynamics. In this sense, the theory is both axiomatic (based on first principles) and mechanistic. In \nameref{Methods} \ref{HarmonicTheory}, we detail the axiomatic derivation, and the mechanistic derivation is detailed in \red{the Supplementary Information, S.2}. The theory is validated using extensive agent-based models, employing two models of spatial decision-making, a spin-system model and a neural-field model \cite{Salahshour2025}, as microscopic realizations of our axioms.  
	
Both routes lead to the same result: behavior can be described as motion on a decision landscape, a structured map that assigns different levels of preference to different possible movement directions (Fig. \ref{Fig0}\textbf{B}). In biological terms, this landscape represents how the organism is inclined to turn in one direction rather than another at a given moment. In physics terms, our approach provides a macroscopic description of behavior by coarse-graining a microscopically detailed one, similarly to how a macroscopic thermodynamic description of inanimate matter can be provided by coarse-graining fast molecular dynamics \cite{Gardiner2004,Zwanzig2001,Zwanzig1960}.
	
The framework rests on four assumptions (axioms). First, sensory inputs are integrated additively, as in standard post-synaptic integration \cite{Dayan2005,Wilson2023,Amari1977,Amit1989,Hopfield1982}. Second, spatial information is represented on a structured neural manifold providing a topological mapping of space \cite{Moser2008,Okeefe1978,Gardner2022,Finkelstein2015,Hafting2005}. In the present study, we use a ring manifold, which is a natural representation for directional choices in two dimensions \cite{Taube2007,Zhang1996,Seelig2015,Kim2017,Pfeiffer2014}. Mathematically, this internal manifold has the topology of the circle ($S^1 \cong \mathbb{R}/2\pi\mathbb{Z}$), so the represented state is an angular variable defined modulo $2\pi$. This is consistent with ring-manifold descriptions of orientation coding in continuous-attractor frameworks \cite{Zhang1996,Seelig2015,Finkelstein2015,Taube2007,Pfeiffer2014,Kim2017,Sarel2017,Mussells2024,Westeinde2024,Wilson2023,Noorman2024}. 
	
These two assumptions, together, can be mathematically formulated by considering the organism to have a ring-shaped map of two-dimensional directional space onto which sensory input is projected, such that a stimulus located at an angular position $\alpha$ induces post-synaptic input on the neurons coding for the corresponding spatial location (see Fig. \ref{Fig0}\textbf{C(i)} and \nameref{Methods} \ref{HarmonicTheory}).
	
Third, the organism’s current decision state (heading direction) is represented by a localized bump of activity on the ring-structured directional manifold \cite{Zhang1996,Kim2017,Seelig2015,TurnerEvans2017,Wilson2023} (see Fig. \ref{Fig0}\textbf{C(ii)}). Mathematically, such a bump is determined by its center, $\phi^{(a)}$, representing the organism's current heading vector, and its width, $W$. Microscopically, it reflects how broadly neural activity underlying heading direction is distributed across the continuous neural manifold representing directional space (commonly thought of as a ring attractor network \cite{Zhang1996,Kim2017,Seelig2015,TurnerEvans2017,Wilson2023}). Macroscopically, $W$ can be interpreted phenomenologically as the uncertainty or volatility of the decision (heading direction). We will thus call it the bump width or decision uncertainty interchangeably. Narrow bumps correspond to precise encoding of directional decisions, whereas wider bumps represent broader and more uncertain internal states.
	
Fourth, neural bump-shape dynamics are assumed to be fast relative to overt behavioral change, so that internal activity settles quickly compared with turning and movement. Under these assumptions, the fast shape variables can be eliminated \cite{Zwanzig1960,Zwanzig2001,Haken1973,Haken1977,Gardiner2004}, leaving a macroscopic description of behavior in terms of macroscopic variables, such as heading direction.
	
A central consequence of this reduction is that the decision landscape can be decomposed into a set of harmonic components---spectral directional patterns that together determine behavior (Fig. \ref{Fig0}\textbf{C}). An individual, therefore, does not react to a target or a neighbor through a single fixed rule. Instead, the nervous system represents directional information as a combination (superposition) of multiple harmonic components, each contributing in its own way to the final movement tendency. Observable behavior reflects the integration of these harmonic components. In this sense, target-seeking, avoidance, choice, and collective coordination all arise from the structure of the same underlying landscape. In physical terms, this landscape can be represented by a Hamiltonian, where the decomposition of the decision landscape into structural components is represented by a summation over harmonic modes, $n$ (Fig. \ref{Fig0}\textbf{D}). Therefore, Harmonic Theory is fundamentally a spectral theory, decomposing the directional change into separate harmonics, or spectral components. Considering a collective of $N$ agents, who perceive each other as sensory input, together with asocial targets, $t$, the Hamiltonian is (\nameref{Methods} \ref{HarmonicTheory}):
	
	\begin{align}
		H_{\mathrm{eff}} = - \sum_{a=1}^{N}\sum_{t} \sum_{n=1}^{\infty} K_n^t(d_{at}) \cos(n(\phi^{(a)} - T^{(a,t)}))- \sum_{a \neq b=1}^{N} \sum_{n=1}^{\infty} K_n(d_{ab}) \cos(n(\phi^{(a)} - T^{(a,b)})).
		\label{eq:HarmonicHamiltonianMain}
	\end{align}
	
Here, $\phi^{(a)}$ is the heading of agent $a$, $T^{(a,t)}$ and $T^{(a,b)}$ are the allocentric bearings from agent $a$ to target $t$ and to agent $b$, respectively, and $d_{at}$ and $d_{ab}$ are the corresponding distances. The first term represents individual-environment interactions (interactions between individuals and inanimate objects), and the second term represents pairwise social interactions. There is no essential distinction between the two (the simplified case of single individual-environment interaction is restored by setting $N=1$), both exhibiting similar harmonic structures, and differences can only arise from the coupling constants, $K_n$ and $K_n^t$ (while both have similar structures). This is the case because, in our framework, social interactions arise when individuals perceive conspecifics as stimuli. 
	
	Notably, while this decision space is derived starting from microscopic neural variables, it is a function of only macroscopic behavioral variables and control parameters, due to the elimination of microscopic variables. In this way, the theory makes the control parameters of behavior transparent: Each harmonic mode contributes with a coupling coefficient which factorizes into a product of three components,
	\begin{align}
		K_n(d_{ab}) = J_{\mathrm{int}}(d_{ab})\, c_n(\sigma)\, M_n(W).
		\label{eq:Kn}
	\end{align}
	
Therefore, the strength of each harmonic component is shaped by three biologically meaningful factors. The first, $J_{\mathrm{int}}(d_{ab})$, is a coupling (possibly distance-dependent) constant which can be thought of as an ecological or environmental interaction kernel. It sets the sign and magnitude of the influence that a target, object, or neighboring individual exerts on the organism, and how that influence changes with distance. Biologically, this term can represent the effective strength of the interaction once sensory information has been projected onto the internal mapping. It can therefore summarize diverse mechanisms, including the decay of stimulus intensity or salience with distance, reduced reliability of social cues from distant neighbors, and, in collective contexts, the switch from long-range attraction to short-range collision avoidance.
	
	We consider several representative cases. (1) For distance-independent attractive stimuli, we take $J_{\mathrm{int}} = h > 0$, corresponding to an attractive target or neighbor. (2) For distance-independent repulsive stimuli, we take $J_{\mathrm{int}} = h < 0$, representing a constant aversive or avoidance-inducing influence. In collective settings, we also investigate two more cases. (3) For distance-dependent interactions, attractive or repulsive, we use an exponential kernel,
	\[
	J_{\mathrm{int}}(d) = h e^{-d/\xi},
	\]
	where $\xi$ is a characteristic interaction range. This should be interpreted as a phenomenological description of influences that weaken with separation, such as attenuation of sensory evidence or decreasing behavioral relevance of distant objects or neighbors. The limit $\xi \to \infty$ recovers distance-independent interactions, whereas smaller $\xi$ (compared to the arena size $L$) produces short-range interactions. (4) We also consider long-range attraction combined with short-range repulsion, with $J_{\mathrm{int}} = h > 0$ beyond a collision radius $r_{\mathrm{coll}}$ and $J_{\mathrm{int}} = h_{\mathrm{coll}} < 0$ for $d < r_{\mathrm{coll}}$. Biologically, this represents organisms that are drawn toward one another at a distance but actively avoid overlap or collision at close range \cite{Couzin2002,Couzin2003,Reynolds1987,McKee2020}.

	The second term of $K_n$ in Eq. \eqref{eq:Kn}, $c_n(\sigma) = \frac{1}{\pi}\exp\!\left(-\frac{n^2\sigma^2}{2}\right)$ is a sensory filter, which describes how precisely directional information is encoded. It arises from the Fourier expansion of the sensory input, which we take to be a Gaussian with width $\sigma$. Large $\sigma$ smooths over fine details, whereas a narrow filter preserves them. Biologically, the width of the sensory kernel, $\sigma$, may depend on environmental conditions, as well as neurobiological factors \cite{Ito2024,Schwartz2011,Carandini2012,Wang2012,Rieke1998} (see Fig. \ref{Fig0}\textbf{C(i)} and \nameref{Methods} \ref{HarmonicTheory}).
	
	For the square bump used here, the third term in Eq. \eqref{eq:Kn}, $M_n(W)=\frac{4}{n\pi}\sin\!\left(\frac{nW}{2}\right)$, is a decision filter, which describes how the organism’s internal decision state integrates incoming information before action is produced (Fig. \ref{Fig0}\textbf{C(ii)}). This term arises from integrating sensory input over the internal neural bump profile and therefore characterizes how the organism’s decision state weights different harmonics. It is an oscillatory function of the bump width, $W$, indicating that changes in neural integration can selectively enhance, suppress, or even reverse particular harmonic contributions, thus modulating the behavioral response to the same stimuli.
	
	Together, these three factors determine which directional components dominate the decision landscape (Fig. \ref{Fig0}\textbf{A}-\textbf{B}). This perspective provides a bridge from mechanism to behavior. The ecological interaction kernel captures what information is available in the world, the sensory filter captures how that information is represented (Fig. \ref{Fig0}\textbf{C(i)}), and the decision filter encapsulates how it is converted into action (Fig. \ref{Fig0}\textbf{C(ii)}). Behavior is therefore not imposed through a catalogue of rules. Rather, it emerges from how sensory information is processed on an internal map of space, leading to a decision landscape prescribing a value to each movement direction (Fig. \ref{Fig0}\textbf{D}). In the following sections, we demonstrate how this single framework accounts for individual target-seeking, avoidance, binary choice, and a range of collective phenomena.
	
	\subsection{The Harmonic Theory of Individual Behavior}
	\subsubsection{Target-seeking}
	
	We begin by studying the simplest case: the response to a single, stationary attractive stimulus. To build intuition for how the Harmonic Theory operates, we first analyze the behavioral consequences of individual harmonic modes in isolation and clarify how they are modulated by the decision uncertainty, $W$ (Fig. \ref{Fig1}\textbf{A}). This step is deliberate; before considering the full superposition of modes, we examine what geometric and dynamical structure each harmonic contributes on its own.
	
	The preferred headings associated with the $n$-th harmonic can be quantified as the minimum-energy solutions of the corresponding harmonic potential at fixed target bearing. This is best seen by transforming the equations of motion in the presence of a target into a coordinate system centered on the target. This leads to a ``tracking error formalism'', developed in \nameref{Methods} \ref{MethodsTargetTracking}, by a change of variable from the heading direction to the ``tracking error’’, defined as the deviation of the agent's heading from the bearing to the target, $\psi=\phi^{(a)}-T^{(a,t)}$. In simple terms, $\psi$ measures how far, in angular space, the agent is from pointing directly at the target. In this formalism, the angular nullclines include both the harmonic steering torque and the change in target bearing due to movement (see \nameref{Methods} \ref{MethodsTargetTracking} and \red{S.4}). 
	
	Examining each harmonic separately reveals a structured hierarchy of behaviors that reflect the symmetry properties of the modes. For $0<W<\pi$, the first two harmonics have $W$-independent preferred headings: The first harmonic favors direct approach towards the target. The second harmonic favors alignment with the target bearing, in which both direct approach and direct escape (aversion to the target) are potential minima. More generally, the $n$-th harmonic has $n$ minima at fixed bearing. Their relation to moving trajectories also depends on distance and kinematic torque (\nameref{Methods} \ref{MethodsTargetTrackingEquilibrium} and \red{S.4}).
	
	This harmonic-by-harmonic analysis clarifies the geometric role of each mode. However, actual behavior arises from the superposition of modes, whose relative strengths are governed by the ecological interaction kernel $J_{\mathrm{int}}=h$, sensory filtering $c_n(\sigma)$, and decision filtering $M_n(W)$. 
	
	When the sensory kernel width $\sigma$ is small, higher harmonics contribute substantially. These modes destabilize the simple approach behavior induced by the first harmonic and generate more complex trajectories, such as spiral motion toward the target intermittently interrupted by direct escape (Fig. \ref{Fig1}\textbf{B}). In purely deterministic dynamics, the agent may fail to reach the target or fail to remain near it due to the dominance of kinematic torques (\nameref{Methods} \ref{MethodsTargetTracking}). Behavioral noise, however, can facilitate both reaching and remaining near the target, particularly when decision-making is governed by higher harmonics (\red{S.4}).

	By contrast, for large $\sigma$, higher harmonics are strongly attenuated, and the first harmonic dominates. The agent exhibits a direct approach to the target and back-and-forth motion upon reaching it (Fig. \ref{Fig1}\textbf{C}). In \nameref{Methods} \ref{MethodsTargetTrackingBifurcation}, we develop a Harmonic Theory of Bifurcations in target-seeking, according to which such back-and-forth motion can be understood as spatial, distance-dependent bifurcations through which direct approach and direct escape alternately lose stability (\nameref{Methods} \ref{MethodsTargetTrackingBifurcation} and \red{S.4}).

	In the \red{Supplementary Information}, we confirm that similar phenomenology, including similar trajectories, also holds in the agent-based neural ring attractor implementation of the theory (\red{S.4}).

	\subsubsection{Response to a stationary aversive target}
	
	Aversive (repelling) behavior arises by reversing the sign of the ecological amplitude $h$. In neural terms, a repulsive stimulus contributes inhibitory rather than excitatory input to the network. For each harmonic $n$, changing the sign of the stimulus leads to a $\pi/n$ shift in fixed-bearing potential minima (\nameref{Methods} \ref{MethodsTargetTrackingTE}). Thus, the fixed-bearing preferred headings rotate; moving trajectories also depend on kinematic torque. 
	
	As in the attractive case, the first two harmonics have $W$-independent preferred headings for $0<W<\pi$: The first harmonic now produces direct escape from the stimulus, and the second harmonic favors headings perpendicular to the target bearing (Fig. \ref{Fig1}\textbf{D}). For higher harmonics ($n>2$), the behavioral outcome depends on $W$. Because $K_n(W)$ can change sign as a function of $W$, higher harmonics may switch between attractive-like and repulsive-like behavior depending on bump width (\nameref{Methods} \ref{MethodsTargetTrackingEquilibrium}). 
	
	This implies a subtle but important point: the same trajectories observed for a repulsive stimulus for a given $W$ may also occur for an attractive stimulus for a different $W$. Higher harmonics alone do not encode whether an interaction is attractive or repulsive; that distinction is determined primarily by the lowest-order modes. Consequently, for small $\sigma$, where higher harmonics contribute more to the trajectory, the agent may move towards a repulsive stimulus and turn away at a late ``decision point'' (Fig. \ref{Fig1}\textbf{E}). By contrast, large $\sigma$, by attenuating the contribution of higher harmonics, leads to early decision in the presence of a repulsive stimulus (Fig. \ref{Fig1}\textbf{F}).

	In the Supplementary Information, we show that, similarly to the attractive stimuli, for repulsive stimuli, noise improves decision-making by facilitating avoidance (\red{S.5}). The agent-based model exhibits similar behavior. See \red{S.5} for details.

	\subsubsection{Choice between two equal targets (binary choice problem)}
	
	We now consider a more complex scenario in which the agent faces two identical attractive stimuli. Empirical data in such settings reveal a bifurcation. At large distances, animals average the two target directions, moving along the symmetry axis between them. As the angular separation between the targets increases, a critical point is reached at which this symmetric trajectory loses stability, and the agent suddenly turns towards, and subsequently selects one of the two targets \cite{Sayin2025,Sridhar2021}.
	
	Assuming the agent begins on the symmetry axis between the targets (Fig. \ref{Fig2}), the harmonic potential (Eq. \eqref{eq:HarmonicHamiltonianMain}) in the presence of two identical targets can be written as (see \nameref{Methods} \ref{MethodsBifurcations}):
	
	\begin{align}
		U(\phi; y) = -2 \sum_{n=1}^{\infty} K_n \cos\left( n \frac{\Delta(y)}{2} \right) \cos\left( n (\phi - \frac{\pi}{2}) \right).
		\label{eq:eqHarmonicPotentialBifurcation}
	\end{align}
	
	Here, $\phi$ denotes the agent's heading direction and $\Delta(y)$ is the egocentric angular separation between the targets when the agent is at $(0,y)$. From Eq. \eqref{eq:eqHarmonicPotentialBifurcation} it is clear that the symmetric, ``averaging trajectory'', $\phi=\pi/2$, is an extremum of the potential for all harmonics (because the first derivative of the potential becomes zero for $\phi=\pi/2$). For each harmonic, the stability of this extremum is determined by the coefficient $K_n \cos\left( n \frac{\Delta(y)}{2} \right)$. 
	
	As the agent approaches the target, $\Delta(y)$ increases, and can change the sign of the cosine term. For the first harmonic, however, the cosine term always remains positive. Thus, the symmetric trajectory remains stable, leading to persistent averaging and a lack of choice. In contrast, for higher harmonics, as the agent approaches the targets, the cosine factor can alternate between positive and negative, making each harmonic alternate between stabilizing and destabilizing forces. Consequently, bifurcations from averaging to choice occur, for each harmonic $n>1$, once the agent reaches a critical angular separation where the coefficient, $K_n \cos\left( n \frac{\Delta(y)}{2} \right)$, crosses zero (Fig. \ref{Fig2}\textbf{A} and \nameref{Methods} \ref{MethodsBifurcations}).
	
	These bifurcations for the first four harmonics are presented in Fig. \ref{Fig2}\textbf{B}, where the stable heading directions as a function of $\Delta$ are plotted. The agent's decision for each scenario (i.e., for given values of $W$ and $\sigma$) results from a weighted superposition of harmonic modes (black lines in Fig. \ref{Fig2}\textbf{B}). For large $\sigma$, the contribution of higher harmonics is attenuated, resulting in a simplified potential landscape where as the agent approaches the target, a stable solution corresponding to averaging (Fig. \ref{Fig2}\textbf{C(i)}) loses stability and gives rise to two symmetric solutions corresponding to choice of one of the targets (Fig. \ref{Fig2}\textbf{C(ii)}). This corresponds to a supercritical pitchfork bifurcation.
	
	In \nameref{Methods} \ref{MethodsBifurcations}, we develop a \textit{Harmonic Theory of Bifurcation} in the binary choice problem by analyzing the harmonic potential Eq. \eqref{eq:eqHarmonicPotentialBifurcation}. The predicted critical angular separation (the smallest angular separation above which the symmetric trajectory loses stability) is presented in Fig. \ref{Fig2}\textbf{D(i)}, and the bifurcation type is presented in Fig. \ref{Fig2}\textbf{D(ii)}. For the fixed square-bump and Gaussian sensory model, numerically resolved bifurcations are supercritical; nearly flat cases require separate consideration (\red{S.6.4}). This frozen-heading classification does not exclude multistability in the full neural or moving-agent dynamics.
	
The bifurcation angle increases with increasing $W$. Intuitively, this can be understood by noting that a bifurcation is expected to occur when the egocentric angular separation of the targets exceeds the bump width, leading to conflicting forces that tend to decompose the bump. Consequently, assuming the bump cannot split, a choice must occur at a critical angular separation. This simple argument thus suggests that the bifurcation angle should be approximately proportional to the bump width---an intuition formalized by the Harmonic Theory.
	
Importantly, Harmonic Theory links the phenomenology, not only during the initial choice stage, but also during the post-decision stage, to the underlying sensory processing structure. The theory provides a link between spatial decision-making and Turing pattern formation in reaction-diffusion systems \cite{Turing1990}. Just like Turing pattern formation, where spatial modes destabilize a homogeneous state, in Harmonic bifurcations, the stabilizing effect of the first harmonic is destabilized by higher harmonics. However, while in Turing pattern formation modes take place in space \cite{Turing1990}, in the Harmonic Theory these modes are defined in the one-dimensional, polar space of heading directions, and their stabilizing or destabilizing nature changes as the agent moves in space, with the frequency of alternation increasing with the harmonic number, $n$.
	
For large $\sigma$, the attenuation of higher harmonics simplifies the potential landscape and thus, trajectories. Upon reaching the targets, the agent often exhibits back-and-forth motion between them (especially for larger $\sigma$), consistent with the Buridan’s paradigm observed in animals such as Drosophila \cite{Colomb2012}, or performs direct escape, with the likelihood of escape increasing as $\sigma$ decreases (Fig. \ref{Fig2}\textbf{E(i)} and \red{S.6.3}). By contrast, for smaller $\sigma$, higher harmonics contribute more strongly and the reported trajectories exhibit richer spatial patterns (Fig. \ref{Fig2}\textbf{E(ii)} and \red{S.6.3}).

In the \red{Supplementary Information}, we show that similar behavior, including comparable pattern formation, emerges in the agent-based neural model (\red{S.6.3}). Furthermore, by considering the effect of noise, we show that the existence of a few solutions for the first and second harmonics makes the low harmonic decision-making resilient to noise, and noise starts to affect the dynamics more strongly for medium harmonics (\red{S.6.5}). For sufficiently high harmonic order, the proliferation of closely spaced solutions produces intrinsically wandering, random-walk-like behavior even in the absence of external noise (see Fig. \ref{Fig2}\textbf{A} for $n=32$ and \red{S.6.2}).
	
Because the agent's behavior is governed by a superposition of harmonics, the effect of noise depends systematically on $\sigma$. For large $\sigma$, higher harmonics are attenuated, and dynamics are dominated by the first few, resilient harmonics; spatial patterns are therefore stable under stochastic perturbations. For small $\sigma$, higher harmonics contribute strongly, generating fine-scale structure that is easily disrupted by noise. Therefore, noise can degrade the detailed spatial patterns observed in the low-noise regime (\red{S.6.5}).
	
In the binary choice problem, noise can play a constructive role at intermediate values of $\sigma$ by facilitating staying close to the targets and reducing the likelihood of leaving them. By contrast, for very large $\sigma$, the agent already remains at the targets deterministically, so noise has little functional effect. For very small $\sigma$, where high harmonics already induce intrinsically irregular dynamics, additional stochasticity does not confer a comparable advantage (\red{S.6.5}). In addition, noise can improve the decision-making speed (leading to a faster choice), especially for large $\sigma$, by facilitating jumping to a decision slightly before the choice becomes the minimum of the potential (\red{S.6.5}). Such noise-assisted choice is reminiscent of stochastic-resonance effects \cite{vanderGroen2018,Trevino2016}.
	
	\subsubsection{Choice between two unequal targets}
	
	We next consider the case of unequal targets. When only the first harmonic is present, the agent forms a weighted average of the target directions biased toward the stronger option (\red{S.7}). The first harmonic alone need not select either target. Introducing the second harmonic can sharpen this preference. The second mode introduces curvature in the decision landscape that favors the stronger target, leading to reliable selection of the better option (\red{S.7}). Consequently, for large $\sigma$, where higher harmonics are attenuated, and dynamics are dominated by the first two modes, the agent exhibits accurate decision-making, strongly preferring the superior target (Fig. \ref{Fig2}\textbf{F(i)}). For smaller $\sigma$, higher harmonics contribute more strongly and can generate bistability in the decision landscape. This bistability can lead to faster (an earlier commitment to a target in some trials) but less accurate decisions (Fig. \ref{Fig2}\textbf{F(ii)}). \red{See S.7 for details}.

	\subsubsection{Response to two aversive targets (binary escape problem)}
	
In the binary escape problem, the agent faces two identical aversive stimuli. In this case, a large difference is observed between the behavior resulting from the first harmonic and higher harmonics: the first harmonic drives direct escape from the average aversive direction. The second harmonic induces motion along an elliptical orbit centered on the average aversive position. Higher harmonics introduce multiple permissible trajectories, many of which include components directed toward the average aversive position. For sufficiently high harmonic order, the proliferation of closely spaced solutions produces irregular, erratic trajectories (Fig. \ref{Fig3}\textbf{A} and \red{S.8}).
	
Notably, Harmonic Theory predicts that similar bifurcations to those in the binary choice problem are present in the binary escape problem. The agent's potential landscape in this case is similar to the former (Eq. \eqref{eq:eqHarmonicPotentialBifurcation}) but with a sign reversal. We present the potential landscape along the symmetric trajectory for two values of the angular separation of the repulsive stimuli, with a large value of $\sigma$, before and after the bifurcation, in Figs. \ref{Fig3}\textbf{B} and \ref{Fig3}\textbf{C}. Before the bifurcation, the potential has a single minimum corresponding to direct escape along the symmetry axis. Closer to the target, the symmetric-approach extremum becomes stable via a subcritical bifurcation (Figs. \ref{Fig3}\textbf{C} and \ref{Fig3}\textbf{D}).

For small $\sigma$, higher harmonics contribute more strongly; the full repulsive landscape can still support competing stable states as the agent moves in space (\red{S.8}). The existence of multiple stable states introduces multistability, such that the agent's trajectory depends on the initial conditions (heading direction), some of which lead to motion towards the average aversive position, or in parallel to the axis joining the two stimuli (Fig. \ref{Fig3}\textbf{E} and \red{S.8}). Notably, noise improves decision-making in the avoidance task, especially for small $\sigma$, where multiple stable states coexist, by facilitating transitions away from approach states (Fig. \ref{Fig3}\textbf{F}). In harmonic terms, noise can reduce the behavioral influence of higher harmonics, which often induce a movement towards the target (see \red{S.8} for details).

In the \red{Supplementary Information}, we show that similar results hold in the agent-based model.

	\subsection{Harmonic Theory of Collective Behavior}
	
	We now scale the theory from individual decisions to collective behavior, where every agent acts as a sensory source for every other agent. 
	
	\subsubsection{Fundamental building blocks of collective behavior: Mechanistic decomposition of collective behavior into harmonic modes}
	
	The effective Hamiltonian for the group is the sum of all pairwise harmonic interactions (Eq. \eqref{eq:HarmonicHamiltonianMain}). A fundamental geometric identity governs these interactions. The bearing from agent $b$ to agent $a$ differs from the bearing from $a$ to $b$ by $\pi$: $T^{(b,a)} = T^{(a,b)} \pm \pi$. This $\pi$-shift acts differently on even and odd harmonics, creating a classification system for behavioral modes into consensus and conflict modes based on a parity principle.

	\paragraph{The parity principle of collective behavior.}
	
	For odd harmonics, $n$, the phase shift is an odd multiple of $\pi$: $\cos(n(x + \pi)) = -\cos(nx)$. The directed terms have opposite signs when expressed relative to the same bearing:
	\begin{align}
		H_{pair}^{(odd)} \propto - \cos(n(\phi^{(a)} - T)) + \cos(n(\phi^{(b)} - T)),
		\label{eq:odd}
	\end{align}
	where $T=T^{(a,b)}$ is the bearing of agent $b$ from $a$. The two agents have phase-shifted preferred headings relative to this common bearing (Figs. \ref{Fig4}\textbf{A(i)} and \ref{Fig4}\textbf{B(i)}). Both can simultaneously minimize the pair potential at distinct headings. Physical exchange also sends $T\to T+\pi$, leaving the pair energy invariant for both parities (\red{S.9}).
	
	By contrast, for even harmonics, the phase shift is an integer multiple of $2\pi$: $\cos(n(x + \pi)) = \cos(nx)$. Consequently, the directed terms have the same sign relative to the common bearing:
	
	\begin{align}
		H_{pair}^{(even)} \propto - \cos(n(\phi^{(a)} - T)) - \cos(n(\phi^{(b)} - T)).
		\label{eq:even}
	\end{align}
	
	This common set of preferred headings can lead to highly ordered collective states, as each pair can simultaneously minimize its energy by adopting the same state relative to the common bearing (Figs. \ref{Fig4}\textbf{C(i)} and \ref{Fig4}\textbf{D(i)}).

	This parity structure provides a mechanistic decomposition of collective behavior into harmonic building blocks, separating collective interactions into bearing-even and bearing-odd classes. The equilibrium heading directions in pairs are derived in \nameref{Methods} \ref{MethodsParityPrinciple} and are presented in Fig. \ref{Fig4}\textbf{A(i)}-\textbf{D(i)} for a static configuration. For pairs, the $n$th harmonic admits $n$ minima per agent, or $n^2$ joint fixed-bearing configurations. For odd harmonics, the bearing reversal results in a $\pi/n$ phase shift in the equilibrium headings of the two agents. When agents are on the move, the changing geometry can generate reconfiguration. By contrast, the bearing symmetry of even harmonics results in identical harmonic potentials, leading to identical sets of preferred heading directions. This can promote consensus, but simultaneous minimization across a population depends on its geometry (\red{S.9}). 
	
	\paragraph{The phenomenology of behavioral modes for odd harmonics.}
	
	To understand how these modes manifest at the group level, we now examine the collective dynamics generated by odd harmonics in pairs and larger groups.
	
	For the first harmonic, the equilibrium configuration under attractive interactions requires $\phi^{(a)}=T$ and $\phi^{(b)}=T+\pi$ (Fig. \ref{Fig4}\textbf{A(i)}). Each agent attempts to orient toward the other’s position. These headings simultaneously minimize the frozen pair potential; movement changes their separation and bearing.
	
	Even in pairs, this mode produces a rich repertoire of behaviors ranging from escape--pursuit to partial synchronization, depending on interaction strength (see Fig. \ref{Fig4}\textbf{A(ii)}, \red{S.9, and SV.1} for example motion patterns). In the small-group examples, changing geometry gives rise to transient patterns, such as rotational milling, vortices, and (unstable) lanes (\red{see S.9 and SV.2 for $N=4$}). In larger groups, the first harmonic is the fundamental force of aggregation, leading to frustrated dense groups (Fig. \ref{Fig4}\textbf{A(iii)} and \red{S.9}).
	
	For aversive interactions, the equilibrium requires $\phi^{(a)}=T+\pi$ and $\phi^{(b)}=T$  (Fig. \ref{Fig4}\textbf{A(iv)}). This role reversal gives rise to evasion and anti-synchronization (Fig. \ref{Fig4}\textbf{A(v)}, \red{S.9 and SV.3}), with the patterns observed depending on the strength of interactions (\red{S.9}). 
	
	The third harmonic is a main generator of complexity in the examples studied. The equilibrium configuration requires: $\phi^{(a)}_k=T^{(a,b)}+2k\pi/3$, for $k \in\{0,1,2\}$ (Fig. \ref{Fig4}\textbf{B(i)}). Complex motion patterns arise from transitions between permissible heading directions, leading to spiraling pairs, weaving trajectories, and sudden direction changes (Fig. \ref{Fig4}\textbf{B(ii)} and \red{S.9 and SV.4}). In larger groups, the multiplicity of this mode leads to fission-fusion dynamics and subgroup formation with rich pattern formation and frequent sudden direction changes resulting in transition between patterns in subgroups (Fig. \ref{Fig4}\textbf{B(iii)}, and \red{S.9 and SV.5}). While similar frustrated and fission-fusion dynamics are observed for higher odd harmonics, the number of their solutions (which is equal to $n$) increases. This multiplicity of solutions leads to a degradation of order in large groups, such that the motion patterns appear to be a disordered motion (see \red{S.9 and SV.8 for an example of motion patterns in groups of $N=80$ for $5$th harmonic}). Consequently, in these examples, the main driver of complex motion patterns is the third harmonic, providing a balance between complexity and order, which is lacking in higher odd harmonics.
	
	For aversive interactions, the motion patterns in pairs for the third harmonic are similar. However, they show a $\pi/3$ phase shift (Fig. \ref{Fig4}\textbf{B(iv)}). This leads to similar spiraling pairs and sudden direction changes, however, with a phase shift (Fig. \ref{Fig4}\textbf{B(v)} and \red{SV.6}). In larger groups, fission-fusion dynamics are observed. Yet, agents keep a higher distance because direct escape is admissible as one of the solutions of all odd harmonics under aversive interactions. Besides, in contrast to attractive interactions where subgroups often form lanes (keeping a zero angular deviation between their heading and group orientation), with aversive interactions, agents maintain a $\pi/3$ heading direction with respect to their common bearing in subgroups (\red{S.9 and SV.7}).
	
	\paragraph{The phenomenology of behavioral modes for even harmonics.}
	
	By contrast to odd harmonics, even harmonics have common preferred headings relative to the pair bearing and can promote consensus and ordered collective states.
	
	The second harmonic is a bearing-axis aligner. The equilibrium is reached for $\phi_1 = T$ and $\phi_2 = T + \pi$ (due to the symmetry of even harmonic potential, we drop the agent’s indices, e.g., $(a)$) (\ref{Fig4}\textbf{C(i)}). Agents align with the axis of the bearing vector but are free to move in either direction along that axis, leading to alignment and antialignment (Fig. \ref{Fig4}\textbf{C(ii)} and \red{S.9}). In collectives, this degeneracy gives rise to bidirectional columns, where subgroups move coherently along a shared axis but in opposite directions  (Fig. \ref{Fig4}\textbf{C(iii)} and \red{S.9 and SV.9}). 
	
	For aversive interactions, admissible  solutions are $\phi_1 = T+\pi/2$ and $\phi_2 = T +3\pi/2$ (\ref{Fig4}\textbf{C(iv)}). In pairs, simulations show alignment perpendicular to the bearing, or stable rotational milling (Fig. \ref{Fig4}\textbf{C(v)} and \red{S.9}). However, because such configurations are difficult (or geometrically impossible for rotational milling) to be satisfied for all pairs in a large group, often subgroups of coherently moving individuals who move perpendicular to their common bearing are observed (\red{S.9 and SV.10 for $N=80$}).
	
	Positive higher even harmonics admit the second-harmonic minima (Fig. \ref{Fig4}\textbf{D(i)}), together with a rich set of new patterns in pairs (Fig. \ref{Fig4}\textbf{D(ii)} and \red{S.9 and SV.11 for an example motion pattern for $n=4$ in pairs}). In large groups, however, due to the increasing multiplicity of solutions, the order resulting from higher harmonics decreases (Fig. \ref{Fig4}\textbf{D(iii)}). This can lead to subgroups of agents who coordinate their motion within themselves, often, while appearing weakly coupled to the rest of the collective (\red{see S.9 and SV.12} for $N=80$ and the fourth harmonic). For aversive interactions with higher even harmonics, a similar decomposition into subgroups is observed, albeit with a different selectivity of motion direction (Fig. \ref{Fig4}\textbf{D(iv)}, \ref{Fig4}\textbf{D(v)}, and \red{SV.13}).

	\subsubsection{Collective behavior resulting from superposition of harmonics by attraction}
	
	Collective behavior emerges from the superposition of harmonic modes, modulated by neuro-ecological parameters through the coefficients $K_n$ (Eq. \eqref{eq:HarmonicHamiltonianMain}).

	\paragraph{Simplification of collective motion by the symmetry of even harmonics and bidirectional column formation.}
	
	Our analysis shows that even harmonics support column formation, a widespread form of collective motion across species. When all the harmonics are superimposed in these examples, the second harmonic---whose solution also holds for higher even harmonics---dominates the motion patterns. However, alignment, arising from the second harmonic alone, does not determine inter-agent spacing (note the high distance between the agents in Fig. \ref{Fig4}\textbf{C(iii)}). The first harmonic reduces inter-agent distance while the second harmonic aligns headings. The interaction of these two effects produces spatially localized, elongated density structures, to which we refer as column formation (Fig. \ref{Fig5}\textbf{B} and \red{S.10.1}). That is, the emergence of dense, elongated clusters of agents that are aligned and extended along the direction of motion (by contrast, bands are elongated structures oriented perpendicular to the direction of travel).

	Because even harmonics admit both alignment and anti-alignment as energetically equivalent solutions under infinite-range interactions, the resulting columns are often bidirectional, with subgroups moving coherently in opposite directions along the same axis (Fig. \ref{Fig5}\textbf{A(i)}), with column width typically decreasing as $W$ decreases. See \red{SV.14} for $W=0.9033$, for an example of a narrow bidirectional column and \red{SV.15} for $W=3\pi/5$, where two separated, sine-wave-shaped but overlapping columns move in opposite directions for larger $W$, and \red{SV.16} for strikingly similar pattern in the neural field agent-based model (see \red{S.10.5}).
	
	\paragraph{Unidirectional column formation via symmetry breaking.}
	
	Harmonic Theory predicts that unidirectional polarized motion cannot arise from even-harmonic symmetry alone. The degeneracy between alignment and anti-alignment must be broken.
	
	In Figs. \ref{Fig5}\textbf{A(ii)} and \ref{Fig5}\textbf{B}, we show that collision avoidance promotes spontaneous selection of a common direction. Increasing the collision radius (below which interactions become effectively negative, favoring avoidance) promotes unidirectional polarization (Fig. \ref{Fig5}\textbf{B}). See \red{SV.17} for the motion patterns in the Harmonic Theory for $W=0.9033$, and \red{SV.18} for $W=3\pi/5$ in the presence of collision avoidance, and \red{SV.19} for similar patterns in the neural-field model.
	
	Similarly, introducing distance-dependent interactions changes density fluctuations and encounters (through the ecological factor $J_{\mathrm{int}}$ in $K_n$). This promotes the formation of unidirectional motion within columns (Fig. \ref{Fig5}\textbf{C}, see \red{S.10.2} and \red{SV.20}). We validate these results using both agent-based models (\red{S.10.5}).
	
	\paragraph{High uncertainty, complex collective motion.}
	
	According to the Harmonic Theory, collective behavior arises from the superposition of harmonics, whose contributions are modulated by the neuro-ecological parameters according to Eq. \eqref{eq:Kn}. Therefore, by changing the weight of different harmonics, a diverse set of collective behaviors is reached by the very same harmonic structure. While for small $W$, the second harmonics dominate, for large $W$, the contribution of the consensus-promoting second harmonics can be attenuated, leading to more complex forms of collective motion. For $W=\pi$, for instance, all even harmonics are eliminated and complex forms of collective behavior, which result exclusively from the complexity-promoting odd harmonics, manifested by strong fission-fusion dynamics, sudden direction changes, and correlated fluctuations, are observed (see \red{S.10.4} and \red{SV.22}). Strikingly similar patterns are observed in the agent-based models (see \red{S.10.5} and compare \red{SV.22} and \red{SV.23}). Here, the disappearance of even harmonics at $W=\pi$ is specific to the square bump (\red{S.2.5}).

	\subsubsection{Collective behavior with only repulsion}

	Because the weight of harmonics can become negative by modulating the intrinsic neural parameter $W$, the Harmonic Theory predicts that similar ordering patterns can be at work when interactions are repulsive, rather than attractive. Therefore, repulsion alone can also lead to a rich suite of collective behaviors, including those with high global order (Fig. \ref{Fig6}\textbf{A} and \red{S.10.3}). These patterns range from ballistic motion in dense subgroups (Fig. \ref{Fig6}\textbf{B} and \red{SV.24}), band formation, where individuals form one, or more parallel bands, moving perpendicular to their common bearing along the band (Fig. \ref{Fig6}\textbf{C}-\textbf{D}, \red{SV.25 and SV.26}), fission-fusion dynamics and correlated fluctuations (Fig. \ref{Fig6}\textbf{E} and \red{SV.27}), rotational milling and subgroup formation, where individuals are decomposed into subgroups, each centred on the vertices of a lattice, performing unidirectional or bidirectional rotational milling (Fig. \ref{Fig6}\textbf{F} and \red{SV.28 to SV.29}). Notably, both band formation and milling patterns are driven by even harmonics, which lead to ``perpendicular alignment'' under repulsive interactions. Consequently, in contrast to the case where ordering is driven by attraction, here the flock orders perpendicular to the band. See \red{S.10.4} for details. We confirm similar phenomenology using the agent-based models (see \red{S.10.5} and \red{SV.30}).

	\subsubsection{Collective motion in bounded space}
	
	Collective motion in both laboratory and natural systems almost always occurs in bounded or structured environments. In experimental studies, animals are typically observed in arenas, tanks, corridors, or enclosures, where spatial boundaries are intrinsic to the setup. Likewise, in natural settings, collective behavior unfolds within landscapes shaped by terrain, obstacles, and resource distributions. Confinement and environmental geometry are therefore fundamental features of the conditions under which collective motion is observed.
	
	By contrast, many theoretical models of collective behavior have focused on effectively unbounded domains implemented via periodic boundary conditions \cite{Vicsek1995,Ginelli2016,Couzin2003,Toner1995,Vicsek2012}. While this abstraction is useful for isolating bulk ordering and phase-transition phenomena, it removes geometric constraints that are central to many empirical contexts \cite{Vicsek2012,Ginelli2016,Ouellette2021,Sumpter2006}. Moreover, minimal local-interaction flocking models often lose cohesion once periodic boundaries are removed, unless additional ingredients---such as long-range attraction or explicit boundary rules---are introduced \cite{Zumaya2018,Armbruster2017,Kuhn2021}.

	Harmonic Theory naturally accommodates bounded domains. Because interactions are formulated geometrically in heading space and superposed across agents, collective structure can emerge and persist under spatial confinement. This makes it possible to directly examine how enclosure geometry shapes the collective modes predicted by the harmonic framework.
	
	In Fig. \ref{Fig7}\textbf{A}, we present the phase diagram and quantitative measures of collective behavior in a bounded domain, specifically, a stadium-shaped arena (Fig. \ref{Fig7}\textbf{B}). Confinement gives rise to a rich set of collective states whose structure depends systematically on the neuro-ecological parameters (\red{S.11.1.1}).
	
	For small $\sigma$, stationary ordering emerges near the walls, reflecting strong higher-harmonic contributions and boundary-induced stabilization (\red{S.11.1.1}). As $\sigma$ increases, a milling phase appears characterized by elevated angular momentum, where the collective closely follows the shape of the arena by turning along the boundaries (\red{SV.31}). For intermediate values of $\sigma$, coherent collective motion with column formation develops (Fig. \ref{Fig7}\textbf{B} and \red{SV.32 and SV.33}). For sufficiently large $\sigma$, aggregation dominates, and global motion weakens.
	
	Increasing the amplitude of interactions $h$ produces additional structured states. These include moving aggregates with ballistic motion (\red{SV.34}), stationary column states exhibiting correlated longitudinal and transverse fluctuations (\red{SV.35}), and fission–fusion dynamics accompanied by the formation of persistent interconnected movement corridors (\red{SV.36}). These structures function as self-organized corridors of motion, reminiscent of transport networks observed in social insects and other animals \cite{Latty2011,PernaLatty2014}. See \red{S.11.1.1} for details.
	
	In addition, confinement itself acts as a dynamical modulator of collective states. Boundary encounters change the real-space geometry, which feeds back into heading-space interactions. Time series of order parameters reveal periodic structure generated by boundary encounters (Fig. \ref{Fig7}\textbf{C}). For example, angular momentum peaks when agents collectively reach and turn along a boundary, whereas polarization is maximal when a coherent column moves through the interior, away from boundary influence.
	
	These observations indicate that collective modes are not determined solely by interaction parameters but also by enclosure geometry. In Fig. \ref{Fig7}\textbf{D}, we further demonstrate this geometric dependence. Depending on the neuro-ecological parameters, different phases, such as column formation (\red{SV.37}) and corridor-network formation (Fig. \ref{Fig7}\textbf{D(i)} and \red{SV.38}) are also observed in a circular arena. However, in a circular arena, consistent boundary curvature can stabilize rotational motion, increasing angular momentum and reducing polarization (Fig. \ref{Fig7}\textbf{D(ii)}, \red{SV.39}, \red{SV.40}, and \red{S.11.2}). Consequently, persistent milling states can be observed in a circular arena. In contrast, in elongated geometries such as the stadium arena, reflection events promote transitions between columnar motion and turning dynamics.
	
	We show in the \red{Supplementary Information, S.11} that increasing neural noise alters quantitative features---such as average distance from boundaries or the strength of order (compare \red{SV.32 and SV.33} for noiseless, and noisy dynamics in collective motion phase)---but does not eliminate the diversity of collective phases. Comparable complexity is observed under stochastic dynamics.

	When only repulsive interactions operate in a bounded domain, collective motion can exhibit strong local and global alignment while remaining largely stationary (\red{S.11.1.2}). In this regime, agents accumulate near the boundaries, and the walls effectively structure the collective configuration. For small $\sigma$, the population separates into two groups, each localized near one end of the stadium (\red{SV.41}). In this configuration, heading directions are only weakly coordinated across the entire group, reflecting the dominance of higher harmonics and local boundary effects.
	
	As $\sigma$ increases, alignment strengthens. Agents distribute more uniformly along the walls, particularly along the long edges of the stadium, adopting similar heading directions. This leads to high global order. However, this high-order state requires the presence of domain walls at the two curved ends of the arena--- transition regions where heading direction changes between oppositely aligned segments (\red{S.11.1.2} and \red{SV.42}).
	
	For larger $\sigma$, alignment becomes sufficiently strong that these domain walls disappear. The system then settles into a globally ordered state in which the group circulates coherently along the boundary, either clockwise or counterclockwise (\red{S.11.1.2} and \red{SV.43}).
	
	Neural noise reduces the stability of these ordered boundary states and can weaken or destabilize global alignment (\red{S.11.1.2}).

	\subsection{A parsimonious theory of spectral, kinematic, and geometrical aspects of behavior}
	
	According to the Harmonic Theory, movement decisions are, fundamentally, an oscillation or relaxation within a potential, which is an explicit sum of harmonics. Thus, the Harmonic Theory is a spectral theory, on the basis of which the temporal evolution of movement decisions, represented in heading direction, $\frac{d\phi}{dt}$, contains important information about the regulation of spatial decision-making. In the \red{Supplementary Information, S.12}, focusing on this aspect of Harmonic Theory and using publicly available data from rummy-nose tetra fish \cite{McKee2020}, we show that the theory exhibits a wide range of alignment with empirical data.
	
	According to the Harmonic Theory, increasing the sensory kernel width attenuates higher angular harmonics. In the simulations examined here, it also reduces the temporal spectral complexity and high-frequency content of heading velocity. Examining several measures of spectral complexity (spectral entropy, centroid, flatness, bandwidth, rolloff, high-power fraction, and peak count), we observe analogous trends with increasing illumination in the fish data (\red{S.12}). These parallel trends do not establish a mapping between illumination and $\sigma$.
	
	This convergence extends to the kinematic and geometric aspects of motion; we observe strong convergence between theory and data in the way polarization in groups is regulated by the same spectral measures, as well as by speed (namely both theory and data predict a decreasing trend between spectral measures and polarization, when polarization is high (\red{S.12})). The comparison also includes descriptive similarities in reported speed and distance-to-wall summaries: both decrease as the sensory kernel decreases in the confined simulations and under lower illumination in the data (\red{S.12}). The reduced equation prescribes a fixed free-space propulsion speed, so the reported speed comparison does not establish intrinsic motor-speed regulation. Finally, by examining the fission-fusion dynamics exhibited in both data and the theory, we show that individuals within subgroups exhibit more similar power spectral densities, indicating shared temporal spectral content (\red{S.12}). 
	
	We do not rule out that some of these convergences may also be exhibited by traditional models. However, the wide range of convergences, encompassing kinematic, spectral, and geometric dimensions of behavior, suggests that Harmonic Theory can provide a unifying, parsimonious explanation for a wide range of observations. This is an especially important point because it suggests that changing environmental conditions need not rewrite the behavioral rules themselves. Instead, environmental change can act by continuously tuning the weights of a shared harmonic basis, thereby moving individuals and groups across behavioral and collective phases.

	\section{Discussion}
	
	Here, we have presented the Harmonic Theory of Behavior, a first-principles framework for spatial decision-making in which individual and collective behavior emerge from how organisms perceive and integrate environmental information on a structured manifold representing directional space. By considering the case of a ring topology, we have demonstrated that the resulting decision landscape naturally decomposes into harmonic components. In this way, Harmonic Theory moves beyond traditional rule-based accounts by deriving the structure of behavior from the geometry of internal representation rather than assuming behavioral rules a priori.
	
	A central conceptual advance is that the theory makes the control parameters of behavior explicit. The effective weight of each harmonic is determined by three biologically interpretable processes: ecological interaction kernel $J_{\mathrm{int}}(d)$, perceptual filtering through the sensory kernel width $\sigma$, and internal neural integration through the bump width $W$. Together, these parameters determine which spatial frequencies most influence behavior and which are suppressed. As a result, changes in sensory conditions, internal state, or interaction range can alter behavior by reorganizing the decision landscape.
	
	\subsection{Individual decisions as spectral bifurcations: alignment with empirical patterns} 
	
	At the individual level, spatial decision-making arises from the superposition of harmonic components that together shape the decision landscape in heading space. Different components contribute to the stabilization of different directional decisions, and behavior reflects which attractors are stable under ecological attenuation, sensory filtering, and internal integration. Sudden changes in direction or sharp choices, therefore, do not require a discrete switch between alternative rules. Instead, they arise when gradual changes in geometry or weighting destabilize one branch of the landscape and stabilize another. When the harmonic components are superimposed, small changes in geometry or these control parameters can reconfigure the stability structure of the landscape, producing the sharp transitions in directions of travel, which may be described as ``choices”. Thus, decision-making arises from parameter-dependent changes, or bifurcations, in the structure of the landscape rather than from switching between discrete rules.
	
	\subsubsection{Single attractive targets}
	
	For a single attractive target, the theory predicts that large $\sigma$ should favor direct target approach, whereas for smaller $\sigma$, higher harmonics can generate curved, spiraling, or delayed-commitment trajectories. This aligns with empirical evidence in animal navigation, such as birds, in which commitment emerges continuously as the geometry of the decision landscape changes along the trajectory \cite{Antolin2023,Schiffner2009}. Similarly, consistent with Harmonic Theory, empirical evidence suggests that target-approach trajectories are less stable and more tortuous when sensory information is weak or unreliable \cite{Newport2021}. While alternative models may result in a similar phenomenology, the relation between degraded sensory precision and the model parameters remains to be established.
	
	The theory further predicts systematic roles for stochasticity. Low-order harmonics are robust because they admit few stable solutions, whereas higher-order structures introduce many competing states that increase sensitivity to fluctuations. In superpositions, noise can either degrade fine-scale structure or stabilize functionality by suppressing higher-order effects, depending on the regime. In particular, the theory predicts that moderate noise can improve weak or ambiguous decisions by suppressing spurious fine-scale structure, in both target-seeking and avoidance tasks. Work on stochastic resonance in perceptual decision-making and visual motion discrimination aligns with this broad finding by showing that stochasticity can improve behavior when sensory evidence is weak \cite{vanderGroen2018,Trevino2016}. 
	
	\subsubsection{Single repulsive targets, predator inspection, and escape geometry}
	
	For a single repulsive target, the theory predicts that the first harmonic favors direct escape, but stronger higher-harmonic contributions can delay commitment to avoidance, or even produce a transient approach followed by a late turn away. This behavior is reminiscent of predator inspection, a counterintuitive tactic in which prey initially approach an apparent predator, but retreat as perceived risk increases \cite{Godin1995,Fishman1999,Veiros2024}. Previous theoretical treatments have often explained predator inspection using cost-benefit and information-gathering frameworks, rather than a simple steering mechanism \cite{DugatkinGodin1992,Fishman1999,YdenbergDill1986,CooperFrederick2007}. 
	
	The Harmonic Theory generates inspection-like trajectories as a deterministic outcome of navigating a multi-harmonic potential landscape under narrow sensory information (small $\sigma$). In this sense, our results suggest a possible mechanistic explanation: when sensory representations are narrow, the model exhibits transient approach even to a repulsive stimulus, producing inspection-like trajectories followed by late avoidance \cite{Godin1995,Fishman1999,Veiros2024}. This is broadly consistent with empirical and conceptual work showing that incomplete or unreliable information can bias antipredator decisions toward assessment and information-gathering before full avoidance \cite{Fishman1999,Crane2024}. Related sensory-ecology experiments show that reduced illumination decreases visual reaction distance \cite{UtnePalm1997}, and behavioral studies indicate that shaded conditions can alter escape tactics toward more cautious responses rather than immediate acceleration or flight \cite{Sabal2021}. While alternative explanations, including information-gathering and game-theoretic accounts, have been proposed \cite{DugatkinGodin1992,Fishman1999,Veiros2024}, according to our theory, inspection-like trajectories can arise from navigation on a repulsive landscape whose higher harmonics remain unsuppressed.
	
	The broader escape literature further shows that prey do not rely on a single stereotyped escape trajectory. Instead, escape often exhibits multiple preferred directions, obstacle-sensitive routing, and geometry-dependent trajectory selection \cite{Domenici2011a,Domenici2011b,Kawabata2023}. This empirical diversity is exactly what one expects from a multi-branch harmonic landscape in which escape reflects structured navigation among competing unstable and metastable directions, rather than a single hard-wired ``move directly away'' rule.

	\subsubsection{Binary and asymmetric choices}
	
	The framework also accounts for why sharp decision transitions occur in continuous heading space. In binary-choice problems between two equal options, the lowest-order component stabilizes averaging, while higher-order components can destabilize this state once geometric separation becomes sufficient. Choice then follows from a loss of stability in the landscape itself. The observed bifurcation is therefore a structural property of the harmonic decomposition rather than a discrete decision module imposed on movement. While the richer predictions of the Harmonic Theory, such as modulation of bifurcation by sensory and environmental factors, remain to be tested in future work, the core phenomenology predicted by the theory has been demonstrated in fruit flies, desert locusts, and larval zebrafish \cite{Sridhar2021,Sayin2025}. In these experiments, individuals initially move along the average of competing options and then undergo a sharp geometry-dependent transition to one branch \cite{Sridhar2021}. The prevalence of this phenomenon across species suggests that the underlying mechanism may be comparatively simple and broadly conserved. While alternative mechanisms, including geometry-dependent conflict between neural populations coding for different targets, have been proposed \cite{Sridhar2021}, the parsimony of the Harmonic Theory and the breadth of its phenomenology make it a candidate explanation. Besides, the theory suggests the same mechanisms can be at work in other decision settings, such as decision-making in the presence of repulsive stimuli.
	
	When the targets are unequal in value or attractiveness, the first harmonic alone gives a weighted average biased toward the stronger target. In the Harmonic Theory, the second harmonic can sharpen this preference and shifts the global minimum toward the superior option. The examples exhibit a speed--accuracy trade-off in the geometry of the decision landscape. A broad sensory kernel (large $\sigma$) suppresses high-frequency components, preserves the low-order bias, and therefore favors slower but more accurate choices. Narrow sensory integration (small $\sigma$) admits more rugged higher-order structures. This can cause earlier bifurcation, but also increases the probability of commitment to an inferior local minimum.
	
	This prediction fits a broad empirical regularity. From bumblebees discriminating among flower options to primates performing saccadic choice tasks, speed--accuracy trade-offs are a pervasive feature of choice behavior \cite{Chittka2003,Heitz2014}. In bumblebees, individuals sacrifice speed for accuracy when errors are penalized \cite{Chittka2003}. Traditional sequential-sampling models, such as the Drift Diffusion Model and the Linear Ballistic Accumulator, typically account for speed--accuracy trade-offs through adjustments of decision boundaries or response thresholds \cite{RatcliffMcKoon2008,BrownHeathcote2008}. However, empirical evidence from humans suggests that speed pressure induces distributed adjustments across sensory, accumulation, and motor stages, rather than being captured solely by a simple lowering of response threshold \cite{Steinemann2018}. The Harmonic Theory demonstrates that the speed--accuracy trade-off does not require a discrete cognitive ``decision threshold'' module. Instead, the trade-off can be a structural consequence of spectral filtering on a neural manifold representing space.
	
	\subsection{Alignment of the Harmonic Theory of Collective Behavior with empirical patterns}
	
	A key result of the theory is that the same spectral logic extends from individuals to groups. Because the vector from one individual to another reverses under role exchange, pairwise harmonic interactions separate into two parity classes. Even modes have common preferred headings relative to the pair bearing, whereas odd modes have phase-shifted preferred headings. Their interplay with group geometry can support coherent order, reorganization, subgrouping, fission--fusion, milling, and persistent rearrangement. The organization of collective behavior is therefore accounted for by a parity principle rather than by an ever-expanding list of interaction rules.
	
	\subsubsection{Bidirectional columns, polarized lanes, and symmetry breaking}
	
	When collective interactions are predominantly attractive, the interplay between the first and second harmonics generates column formation. When interactions are sufficiently long-ranged, the symmetry of the second harmonic can produce bidirectional columns in the model, in which subgroups move coherently along a shared axis but in opposite directions, similar to bidirectional traffic, for instance, in social insects \cite{PernaLatty2014,Latty2011,CouzinFranks2003,Wang2018}. In this framework, such states reflect the degeneracy of the second-harmonic landscape. 
	
	The theory predicts that robust unidirectional polarized motion requires symmetry breaking. In the model, spontaneous direction selection is promoted by collision avoidance or distance-dependent interactions, although radial interactions preserve the even-harmonic heading-reversal symmetry (\red{S.9}). This may help interpret why unidirectional polarized motion is more common in systems with stronger short-range repulsion or more local interactions, such as migratory locusts \cite{Bazazi2008,Sayin2025} and schooling fish \cite{Katz2011,Herbert-Read2011}.
	
	\subsubsection{Aversion as a source of order rather than disorder}
	
	One of the surprising and counterintuitive predictions of the Harmonic Theory is that aversive interactions (repulsion) need not destroy order. Instead, it can promote diverse forms of order, from band formation to milling. While it remains unclear whether repulsive interactions alone generally underlie ordered collective motion in animals, short-range repulsive interactions are an important component of the inferred interaction rules in several species, including schooling fish \cite{Herbert-Read2011,Calovi2018}. In locusts, frontal collision avoidance and anisotropic neighbour interactions coexist with large-scale migratory order, playing an essential role in the formation of migratory bands \cite{Bazazi2008,Weinburd2024}.
	
	\subsubsection{Regulation of collective motion by sensory and environmental factors}
	
	The Harmonic Theory predicts that changes in sensory range or interaction range should reshape collective states by reweighting the harmonic spectrum. Experiments on rummy-nose tetra fish show that schooling depends critically on vision: groups require a minimum illuminance to achieve high polarization, whereas compromising visual information weakens schooling \cite{McKee2020}. More generally, illumination modulates the strength and effective range of social interactions in schooling fish \cite{Lafoux2023,Xue2023,Lombana2022}. Furthermore, light modulates several aspects of collective motion, from geometric to spectral and kinematic aspects, with qualitative parallels in the theory; this does not establish an illumination--$\sigma$ relation.
	
Our analyses also suggest that the harmonic picture is not limited to static spatial order parameters. In the fish data, increased polarization and decreased inter-individual distance are associated with greater similarity of heading-velocity spectra, whereas lower light is associated with stronger wall use, lower speed, and more complex spectral content (\red{S.12}). These results suggest that harmonic weighting can reorganize both spatial order and temporal motion. Angular harmonic order and temporal frequency are distinct, and power-spectrum similarity alone does not establish synchronization.
	
	The framework also naturally incorporates bounded environments by showing that confinement reshapes which collective modes are expressed through reflection and curvature. These predictions are broadly consistent with empirical observations that enclosure geometry and wall interactions can organize collective motion \cite{Lecheval2018,Calovi2018,Xue2023}.

	\subsection{Broader implications, limitations, and outlook}
	
	\subsubsection{Link to normative views of decision-making}
	
	Harmonic Theory can also provide a mechanistic foundation for top-down normative approaches to decision-making \cite{Summerfield2022,Barendregt2022,Kilpatrick2019}. Normative models often describe behavior in terms of objective functions, adaptive decision rules, or optimal policies \cite{Summerfield2022,Barendregt2022,Kilpatrick2019}. In the Harmonic Theory, however, the relevant objective is not postulated a priori; rather, it emerges as an effective potential governing behavior through stochastic dynamics on a decision landscape, obtained via bottom-up coarse-graining in which fast neural dynamics are reduced to an effective Hamiltonian \cite{Gardiner2004,Pathria2017}. Because this Hamiltonian depends on a small set of biologically interpretable parameters, it provides a mechanistically grounded quantity that plays a role analogous to an objective function that can, in principle, be fitted to behavioral data. In this way, the theory bridges causal explanation and normative inference, allowing top-down descriptions to be anchored in sensory and neural mechanisms rather than treated as purely phenomenological fitting forms. In this respect, the framework is analogous in spirit to the historical development of classical mechanics, where variational principles (such as Hamiltonian formalism) supplied compact macroscopic descriptions of motion anchored in underlying dynamical laws \cite{Goldstein1950}. 
	
	\subsubsection{Limitations and future work}
	
	While the evidence discussed here supports the geometric logic of the theory, comprehensive quantitative tests of its predictions are still lacking. The framework makes explicit predictions about how changes in sensory kernel width, interaction range, and internal integration should reshape the spatial and temporal spectra of behavior. These predictions provide clear targets for future empirical studies.

	At a broader conceptual level, the present work shows that environmental information processing employing a topological mapping of space can give rise to the Harmonic Theory of Behavior. The ring case is a tractable and biologically widespread starting point for directional behavior, not a statement that all cognition is one-dimensional. More generally, however, the underlying logic of the framework is not restricted to ring topology. Because the derivation begins from spatial information processing on a topological mapping, the same bottom-up program could, in future work, be extended to other topologies of neural representation, including cases in which space is encoded on manifolds with different symmetries and dimensions.

	\subsection{Conclusion}
	
	In summary, the study of collective behavior in living systems has drawn heavily on statistical physics and self-organization approaches to explain how macroscopic order emerges from local interactions. The Harmonic Theory builds on this foundation, but shifts the focus from interaction rules to the structure of internal representations and the processing of spatial information from which behavior emerges. The theory arises from biologically grounded principles underlying the transformation of environmental cues into movement decisions employing a topological mapping of space. By coarse-graining microscopic neural dynamics, in the spirit of non-equilibrium statistical mechanics \cite{Zwanzig2001,Kardar2007,Pathria2017,Marchetti2013}, the theory provides a mechanistic bridge from perceptual processing and neural integration to observable behavior in individuals and collectives. It replaces a catalogue of behavioral rules with a small set of geometric components whose relative weighting can be predicted from ecological attenuation, perceptual filtering, and internal integration. This unified structure generates testable predictions for how changes in perceptual conditions, internal state, and environmental geometry should reshape both decision landscapes and emergent individual and collective behavior.

	\section{Methods}
	\label{Methods}
	
	\subsection{The Harmonic Theory of Behavior}
	\label{HarmonicTheory}
	To ground the theory in first principles, we derive the effective macroscopic theory from a set of minimal, biologically plausible assumptions regarding the agent's sensory-motor organization. This approach ensures the generality of the framework, making it applicable, under the assumptions below, to an organism (or robot) that processes spatial information via a ring-structured topological mapping of directional space \cite{Dayan2005,Moser2008,Palgi2025,Okeefe1978}. To do so, we show that any system satisfying a few general assumptions---linear environmental coupling, a ring-like topological map, and time-scale separation---will obey the formalism of the Harmonic Theory. We validate our results using two specific realizations: a spin-system model and a neural-field model introduced before \cite{Salahshour2025}.
	
	\subsubsection{A first-principles approach}
	
	\textbf{Assumption 1: Linear coupling with the environment.} Our most fundamental assumption is the linearity of coupling between the neural system and the environment. Let $s(\alpha)$ denote the state of a neural population (e.g., the firing rate or membrane potential) responding to sensory input localized at a spatial state $\alpha$, and $h(\alpha)$ the environmental input (sensory field) on such neurons. Mathematically, the linearity assumption can be applied by positing that the interaction energy resulting from the coupling of the neural system with the environment, $H_{env}$, is the overlap (correlation) between the internal neural state and the external sensory field:
	$$H_{env}(\phi) \propto - \int s(\alpha) h(\alpha) d\alpha$$
	This linearity represents the summation of post-synaptic currents \cite{Dayan2005,Wilson1972,Amit1989}, a common feature of neural network models \cite{Hopfield1982,Amari1977,Dayan2005}. In biological terms, it implies that the directional preference for a stimulus depends on how well its internal representation $s(\alpha)$ overlaps with the external input $h(\alpha)$. The minus sign indicates that the system seeks to minimize energy by maximizing this overlap---i.e., favoring overlap with attractive input.

	The ``environmental state'' $\alpha$ determines how the space is represented by the neural system. This is shaped by the network topology. Empirical evidence has pointed to different ways that space can be represented in animal (and human) brains, such as grid-like \cite{Moser2008,Hafting2005}, toroidal \cite{Gardner2022}, or ring-like topological maps of space \cite{Kim2017,Taube2007}. While an extension of our framework to other topologies is, in principle, possible, here, we limit ourselves to ring topological representation, a common neural motif employed across levels of organization to represent spatial directions \cite{Kim2017,Seelig2015,Sarel2017,Finkelstein2015}.

	\textbf{Assumption 2: The sensory ring.} We assume that the agent's sensory neurons are topologically arranged on a ring ($S^1 \cong \mathbb{R}/2\pi\mathbb{Z}$), representing the agent’s allocentric (world-centered) angular space $\alpha \in [0, 2\pi)$, leading to the sensory coupling, $H_{env} = - \frac{N_s}{2\pi}\int_0^{2\pi} s(\alpha) h(\alpha) d\alpha$.
	Here, we have introduced the prefactor $\frac{N_s}{2\pi}$, in anticipation of a discrete, mechanistic implementation of our approach. This assumption is biologically grounded in the architecture of the central complex in insects \cite{Kim2017,Seelig2015}, and the neural architecture underlying the establishment of head direction \cite{Zhang1996,Seelig2015,Finkelstein2015}, heading direction \cite{Taube2007,Pfeiffer2014,Kim2017} and goal direction \cite{Sarel2017,Mussells2024,Westeinde2024,Wilson2023}, in invertebrates and vertebrates, where neurons are arranged on a topographical ring to encode directions \cite{Taube2007,Zhang1996,Kim2017}.

	We model the sensory input $h(\alpha)$ produced by an external object at bearing $T_{\mathrm{target}}$ as a distributed field. Biologically, this reflects that a localized object can produce an angular sensitivity profile, and downstream pooling further smooths the representation \cite{EnrothCugell1966,Dumoulin2008}. In flies, for instance, visual signals reach the central complex through identified pathways from the optic lobe to the anterior optic tubercle and bulb and onward to ellipsoid-body ring neurons; these inputs in turn modulate the heading/landmark representation that is topographically mapped across central-complex compartments, including the protocerebral bridge \cite{Seelig2015,Lovick2017,Omoto2017,Chang2017}. In this sense, each object can be treated computationally as an external field acting on the neural network, providing a mapping of space. In insects and flies, such angular sensitivity functions are commonly well-approximated by Gaussian profiles \cite{Stavenga2003}. Furthermore, Gaussian profiles (and sums/differences of Gaussians) are a standard, effective description of spatial sensitivity fall-off in early vision and of population receptive-field models that summarize the net spatial pooling of neural populations \cite{EnrothCugell1966,Dumoulin2008}. We therefore model the object(s) as inducing a localized input:
	\[
	h(\alpha)=\sum_{\mathrm{target}}\frac{J_{\mathrm{int}}^{0}}{\sqrt{2\pi\sigma^2}}\exp\!\left(-\frac{(\alpha-T_{\mathrm{target}})^2}{2\sigma^2}\right),
	\]
	where $J_{\mathrm{int}}^{0}$ sets microscopic stimulus strength and the sensory kernel, $\sigma$, is the effective angular integration width of the sensory pathway projecting onto the ring. 
	
	Biologically, $\sigma$ may represent the width of the post-threshold drive that reaches the circuit after early nonlinear processing (filtering, rectification/thresholding, and gain control/normalization) \cite{Schwartz2011,Carandini2012,Wang2012}. Under dim illumination, photon-shot noise and intrinsic dark noise reduce the signal-to-noise ratio, and retinal/early-visual pathways employ nonlinear thresholds that suppress weak inputs and preserve only the most reliable components \cite{Rieke1998,Schwartz2011}. In such a cascade, the tails of a broad optical blur can fall below threshold, so that the suprathreshold angular profile that actually drives the ring is narrower. This is a possible mechanism, but it does not establish how $\sigma$ depends on illumination. More generally, $\sigma$ can also subsume other factors that change the angular footprint of evidence, including target angular size/distance, stimulus contrast and turbidity (visibility), as well as neurobiological factors \cite{Ito2024}.

	Experiments and data-driven models in schooling fish demonstrate that illuminance modulates the strength and range of social interactions \cite{Lafoux2023,Xue2023,Lombana2022}. These effects do not identify $\sigma$, which describes the angular width of each cue on the ring, rather than the extent of the usable visual field.
	
	Under assumptions 1 and 2, the energy function (Hamiltonian) of the neural system, governing agent $a$ can be written as $H^{(a)}=H_0^{(a)}+H_{env}^{(a)}$, which for a collective of $N$ agents is straightforwardly extended to $H=\sum_{a=1}^{N}H^{(a)}=\sum_{a=1}^{N}(H_0^{(a)}+H_{env}^{(a)})$. Further progress requires specification of the neural system, i.e., $H_0$, which depends on the microscopic realization of the network. However, importantly, the coupling between the neural dynamics and the environment is given by the interaction term, $H_{env}$. Consequently, a macroscopic theory with wide applicability can be derived by focusing on this term and making two further, broad and empirically motivated assumptions which keep our theory consistent with a wide range of microscopic realizations.
	
	\textbf{Assumption 3: The Bump Ansatz (Decision State).}
	We assume that the intrinsic dynamics of the neural network compel the system to settle into a stable, localized ``bump" centered on an angle, $\phi^{(a)}$, and with width $W$. As shown in foundational works in neural-field models, this assumption is a broad property of attractor neural networks \cite{Amari1977}, where, depending on the network at hand, the localized bump, $\phi^{(a)}$, can represent the head direction \cite{Zhang1996,Seelig2015,Finkelstein2015}, heading direction \cite{Taube2007,Pfeiffer2014,Kim2017} and goal direction \cite{Sarel2017,Mussells2024,Westeinde2024,Wilson2023}. In our framework, $\phi^{(a)}$ represents the agent's emergent heading direction, which we take to be the same as the agent's goal direction, assuming that the agent can turn toward its goal direction \cite{Salahshour2025}. Mathematically, taking the bump to have a square shape, assumption 3 states:
	$$
	s^{(a)}(\alpha;\phi^{(a)},W^{(a)}) = 
	\begin{cases} 
		+1 & \text{if } \alpha \in [\phi^{(a)} - W^{(a)}/2, \phi^{(a)} + W^{(a)}/2] \\ 
		-1 & \text{otherwise}
	\end{cases}
	$$
	We note that the width, $W$, can be interpreted phenomenologically as the decision uncertainty. A narrow bump ($W^{(a)} \to 0$) implies a precise, high-confidence decision. A wide bump implies a diffuse, uncertain decision.
	
	\textbf{Assumption 4: Time scale separation (adiabatic approximation).} A foundational step in deriving an emergent theory of behavior is the separation of timescales. Neural dynamics occur on the order of milliseconds \cite{Sudhof2013,AbbottRegehr2004}, while behavioral adjustments occur on the order of seconds \cite{Lafoux2023}. Consequently, we assume an adiabatic separation of timescales \cite{Gardiner1984,VanKampen1985,Gardiner2004}. The neural bump shape rapidly relaxes at a given bump center before the agent physically turns. This assumption is equivalent to the Haken’s Slaving Principle in synergetics \cite{Haken1977,Haken1973}, according to which, the ``order parameters" (slow variables) determine the behavior of the individual parts (fast variables), which essentially follow the order parameters instantaneously. This allows us to ``integrate out" the fast microscopic neural fluctuations, treating the equilibrium neural configuration (which we take to be a bump according to assumption 3) as a stable macroscopic object that guides the slower behavioral variable, the heading $\phi^{(a)}$.
	
	Mathematically, assumptions 3 and 4 allow us to write the interaction energy as a function of the macroscopic variables: 
	\begin{align}
		H_{env}^{(a)}(\phi^{(a)},W^{(a)}) = - \frac{N_s}{2\pi}\int_0^{2\pi} s(\alpha;\phi^{(a)},W^{(a)}) h(\alpha) d\alpha.
		\label{eq:HamiltonianMethods}
	\end{align}

	\subsubsection{Effective Hamiltonian and the ``equations of motion'' of the macroscopic behavior}
	
	For a fixed translated bump profile, assumptions 3 and 4 allow us to derive the macroscopic, effective Hamiltonian from the sensory overlap (\red{S.2.5}). The resulting theory is applicable to a broad range of networks that are consistent with our assumptions. We will, shortly, provide two such microscopic realizations.
	
	The core mathematical operation in the Harmonic Theory is the derivation of the effective Hamiltonian $H_{\mathrm{eff}}$ by computing the overlap integral $H_{env}^{(a)}$, using the spectral properties of the sensory and neural fields. We expand the periodic Gaussian sensory input $h(\alpha)$ into a Fourier cosine series (\red{S.2}):
	\begin{align}
		\sum_{k\in\mathbb Z}\frac{1}{\sqrt{2\pi\sigma^2}} \exp\big[-\frac{(\alpha_i^a-T^{(a,b)}+2\pi k)^2}{2\sigma^2}\big] = \sum_{n=0}^{\infty} c_n \cos(n(\alpha_i^a - T^{(a,b)})),
		\label{eq:FourierMethods}
	\end{align}
	where $c_0=1/(2\pi)$ and, for $n\geq1$, $c_n(\sigma) = \frac{1}{\pi} \exp\left(-\frac{n^2 \sigma^2}{2}\right)$. These coefficients are exact for the wrapped Gaussian; for a Gaussian of shortest angular distance they are a narrow-kernel approximation (\red{S.2.5}).
	Using Eq. \eqref{eq:FourierMethods} in Eq. \eqref{eq:HamiltonianMethods}, the interaction Hamiltonian becomes:
	\begin{align}
		H_{env}^{(a)}=-\sum_{b\neq a}\sum_{n=1}^{\infty}\frac{N_s}{2\pi} J_{\mathrm{int}}^{0}(d_{ab}) c_n\int_0^{2\pi} s^{(a)}(\alpha) \cos(n(\alpha - T^{(a,b)})) d\alpha. 
	\end{align}
	Here, the first summation originates from considering each agent is a target for every other agent (an extension to asocial targets is straightforward). Using assumption 3 for the neural state, this integral can be solved, giving rise to the effective, macroscopic Hamiltonian in which the microscopic neural variables are eliminated. Therefore, the effective Hamiltonian is a function of macroscopic behavioral variables, $\{\phi^{(a)}\}$, and neuro-ecological parameters, neural bump width, $\{W^{(a)}\}$, interaction kernel $J_{\mathrm{int}}$, allocentric bearings, $\{T^{(a,b)}\}$, and distance between agents (and/or between agent(s) and target(s)), $\{d_{ab}\}$:
	
	\begin{align}
		H_{env} = - \sum_{a \neq b} \sum_{n=1}^{\infty} K_n(d_{ab}) \cos(n(\phi^{(a)} - T^{(a,b)})),
		\label{eq:HarmonicHamiltonian}
	\end{align}
	
	where:
	\begin{align}
		K_n(d_{ab}) = J_{\mathrm{int}}(d_{ab}) \cdot c_n \cdot M_n^{(a)}.
	\end{align}
	Here, $J_{\mathrm{int}}=(N_s/2)J_{\mathrm{int}}^{0}$ absorbs the neural density factor, with $J_{\mathrm{int}}^{0}$ the microscopic input amplitude. Below, we clarify the meaning of these terms (\red{S.2}).
	\begin{itemize}
		\item \textbf{$J_{\mathrm{int}}(d)$} is a coupling constant which we call the ecological interaction kernel. We will examine different cases, where the interaction kernel is a constant (independent of distance), $h$, or depends on distance, and when it is positive (attractive stimuli), or negative (repulsive stimuli). In the general case, where we consider short-range repulsion and long-range, distance-dependent attraction, this is taken to be:
		$$ J_{\mathrm{int}}(d) = \begin{cases} -|h_{\mathrm{coll}}| & \text{if } d \le r_{\mathrm{coll}}, \\ +h \exp(-d/\xi) & \text{if } d > r_{\mathrm{coll}}. \end{cases} $$
		Here, $\xi$ is a characteristic length scale of interactions, beyond which the strength of the stimuli falls rapidly.
		
		\item \textbf{$c_n$} is the Fourier coefficient of the sensory input, which depends on the receptive field width $\sigma$:
		$$c_n = \frac{1}{\pi} \exp\left(-\frac{n^2 \sigma^2}{2}\right)$$
		\item \textbf{$M_n$} is the n-th order moment of the neural bump, which depends on the stable bump width $W$:
		$$M_n = \frac{4}{n\pi} \sin\left(\frac{nW}{2}\right)$$
	\end{itemize}
	Eq. \eqref{eq:HarmonicHamiltonian} encapsulates the foundational ``laws'' of the Harmonic Theory. It states that behavioral decisions are driven by a superposition of harmonic modes. The agent does not respond to the raw position of the target; rather, its heading couples to the $n$-th harmonic of the bearing $T^{(a,b)}$ with a strength determined by the coupling constant $K_n$.
	
	The complete effective Hamiltonian is $H_{\mathrm{eff}}^{(a)} = H_{0}^{(a)} + H_{\mathrm{env}}^{(a)}$. However, if the recurrent network is homogeneous along the ring (rotationally symmetric) \cite{Amari1977,Zhang1996,Kim2017}, then $H_{0}^{(a)}$ is invariant under shifts of the bump center and therefore contributes only an additive constant (independent of $\phi^{(a)}$) to $H_{\mathrm{eff}}^{(a)}$. Since $H_{\mathrm{eff}}^{(a)}$ enters the equations of motion only through its derivatives with respect to $\phi^{(a)}$, this constant can be dropped without loss of generality \cite{Kardar2007,Pathria2017}. We therefore take the effective Hamiltonian to be the part induced by interactions with the environment, $H_{\mathrm{env}}^{(a)}$. Following the standard language of stochastic processes and non-equilibrium statistical physics, the dynamics of the agent's observable heading $\phi^{(a)}$ (and thus position) are modeled by an overdamped gradient Langevin equation driven by the effective Hamiltonian (\red{S.2.5}) \cite{Gardiner2004,Zwanzig2001,Kardar2007}:
	\begin{align}
		\dot{\vec{r}}^{(a)} &= v_0^{\mathrm{eff}} \begin{pmatrix} \cos(\phi^{(a)}) \\ \sin(\phi^{(a)}) \end{pmatrix} \nonumber\\
		\dot{\phi}^{(a)} &= -\eta \frac{\partial H_{\mathrm{eff}}^{(a)}}{\partial \phi^{(a)}} + \sqrt{2D_r} \epsilon(t).
		\label{eq:motion}
	\end{align}
	Here, $\eta$ is the rotational mobility (responsiveness).
	$D_r$ is the rotational diffusion coefficient.
	$\epsilon(t)$ is Gaussian white noise.
	The macroscopic noise strength is related to the microscopic neural noise, phenomenologically, via a fluctuation-dissipation relation \cite{Kardar2007,Gardiner2004}: $D_r/\eta \propto 1/\beta$. This phenomenological link allows us to predict how neural noise (e.g., variability in spike timing) propagates up to observable behavior. See \red{S.1 and S.2} for details.
	
	\subsubsection{A microscopic realization: the spin-system model}
	
	Above, we derived the Harmonic Theory using a first-principles approach. However, this theory can also be derived starting from a microscopic realization, in which case, the effective parameters, $W^{(a)}$ and $v_0^{\mathrm{eff}}$, are linked to the underlying neural variables. While we validate the theory using two realizations, demonstrating that our four assumptions are naturally satisfied by standard neural architectures, in the \red{Supplementary Information} we mathematically derive the theory using a spin-system model which we have introduced before \cite{Salahshour2025}.
	
	In this model, which is consistent with assumptions 1, 2, and 4 (and assumption 3 arises as an emergent property), the neural network is modeled as a ring of $N_s$ Ising spins $s_i \in \{+1, -1\}$. The Hamiltonian representing the internal neural dynamics, $H_0$ for a single agent includes recurrent connectivity $J_{ij}$ and global inhibition $h_b$, $H_0^{}(a)=- \left[ \frac{1}{N_s} \sum_{i,j} J_{ij} s_i s_j - h_b \sum_i s_i \right]$, and the interaction term is given by $H_{env}^{(a)}=-\sum_{i}s_ih_i$, giving rise to:
	
	\begin{align}
		H^{(a)} =H_0^{(a)}+H_{env}^{(a)}= - \left[ \frac{1}{N_s} \sum_{i,j} J_{ij} s_i s_j + \sum_i h_i s_i - h_b \sum_i s_i \right].
		\label{eq:spinsystemModel}
	\end{align}
	The neural system operates at a ``neural temperature'' $1/\beta$, which characterizes the level of stochastic noise in neural dynamics \cite{Kardar2007,Pathria2017}. High $\beta$ implies low noise (deterministic, precise decision-making), while low $\beta$ implies high noise (stochastic, exploratory behavior). For the network kernel, $J_{ij}$, we employ a generalized cosine interaction profile $J_{ij} = \cos(\pi|(\alpha_i - \alpha_j)/\pi|^\nu)$, which promotes the formation of a stable activity bump (thus, assumption 3 mechanistically emerges) \cite{Salahshour2025}. Finally, we assume the agent moves according to the profile of its active neurons such that the speed vector is given by $\vec{v}=\frac{v_0}{N_s}\sum_{i=1}^{N_s}\vec{\alpha}_i\delta_{s_i,+1}$ \cite{Salahshour2025}. 
	
	Using a simple cosine-shaped synaptic kernel, $J_{ij} = \cos(|\alpha_i - \alpha_j|)$, in the \red{Supplementary Information}, and using the bump ansatz (which is an accurate assumption in the high-$\beta$, low-noise regime of neural dynamics \cite{Salahshour2025}), we show that the stable bump width $W$ is controlled by the global inhibition $h_b$ via the relation $\sin(W) = \frac{\pi h_b}{2}$ (on the locally stable branch; \red{S.2}) and the agent's speed constant is given by $v_0^{\mathrm{eff}}=v_0\frac{1}{\pi}\sin(W/2)$. Thus, these effective parameters are mechanistically linked to the physiological parameters of the specific neural circuit.

	\subsection{The agent-based models}
	\label{MethodsABMs}
	In the Supplementary Information, we validate the theory using two agent-based models, the spin-system model (described above) and the neural-field model \cite{Salahshour2025}. Despite their differences in the implementation of the neural dynamics, these models share the core assumptions of the Harmonic Theory: namely, sensory integration via post-synaptic input. The spin-system model is defined by the Hamiltonian given in Eq. \eqref{eq:spinsystemModel}. The neural-field model incorporates similar terms; however, in this model, the neural dynamics obey an Amari-style integrodifferential equation. See \red{S.3} for details.

	\subsection{Target-seeking and single-target bifurcations in the Harmonic Theory}
	\label{MethodsTargetTracking}
	In this section, we provide a mathematical analysis of target-seeking behavior, in which a single agent faces a single target.
	\subsubsection{The geometrical representation of harmonics}
	\label{MethodsTargetTrackingEquilibrium}
	For attractive stimuli, the minimum of the harmonic energy Eq. \eqref{eq:HarmonicHamiltonianMain} is reached when $\cos(n(\phi^{(a)} - T^{(a,t)}))=\sgn(\sin(nW/2))$, leading to a ``tracking error'' (deviation of the heading direction from the allocentric bearing to the target):
	\begin{align}
		\psi=
		\begin{cases}
			\phi^{(a)}-T^{(a,t)}=2k\pi/n$, \quad if\quad  $\sgn(\sin(nW/2))>0\\
			\phi^{(a)}-T^{(a,t)}=(2k+1)\pi/n$ \quad if \quad $\sgn(\sin(nW/2))<0,
		\end{cases}
		\text{for}\quad k\in\{1,\ldots,n\}.
	\end{align} 
	where the solutions are modulo $2\pi$. For $0<W<\pi$, the preferred headings of the first two harmonics are independent of $W$: direct approach, $\psi=0$, for the first harmonic, and alignment with the target, where both the direct approach $\psi=0$, and direct escape $\psi=\pi$, are potential minima, for the second harmonic (Fig. \ref{Fig1}\textbf{A}). For $\pi<W<2\pi$, the second harmonic changes sign. Each nonzero harmonic, $n$, admits $n$ fixed-bearing minima at tracking errors $2k\pi/n$ or $(2k+1)\pi/n$ (depending on $\sgn(\sin(nW/2))$), with the target. 
	
	For odd harmonics, the fixed-bearing minima include either direct approach or direct escape, depending on the sign of $K_n$, and additional oblique headings for $n>1$.
	
	For even harmonics, direct approach and escape are either both minima or neither is a minimum. Perpendicular minima do not imply circular trajectories: a pure even-harmonic torque vanishes at $\psi=\pm\pi/2$, whereas circular motion requires nonzero turning (\red{S.4}).
	
	For repulsive stimuli, the minimum of the harmonic energy is reached when $\cos(n(\phi^{(a)} - T^{(a,t)}))=-\sgn(\sin(nW/2))$, leading to a tracking error:
	\begin{align}
		\psi=
		\begin{cases}
			\phi^{(a)}-T^{(a,t)}=(2k+1)\pi/n$, \text{ if } $\sgn(\sin(nW/2))>0\\
			\phi^{(a)}-T^{(a,t)}=(2k)\pi/n$, \text{ if } $\sgn(\sin(nW/2))<0,
		\end{cases}
		\text{for}\quad k\in\{1,\ldots,n\}.
	\end{align}
	Here, for $0<W<\pi$, the preferred headings are direct escape, $\psi=\pi$, for the first harmonic, and perpendicular headings, $\psi=\pi/2$ or $3\pi/2$, for the second harmonic. These are fixed-bearing minima, not prescribed moving trajectories.
	
	This decomposition clarifies that the preferred headings of a single harmonic with $n>1$ do not uniquely identify attractive versus repulsive stimuli when $W$ is unknown due to the possibility of sign change in the potential depending on $W$: the fixed-bearing minima of the $n$th harmonic for attractive stimuli for values of $W$ satisfying $\sgn(\sin(nW/2))>0$, would be identical to those observed for a repulsive stimuli, but for different values of $W$, satisfying $\sgn(\sin(nW/2))<0$.

	\subsubsection{The tracking error coordinate in target-seeking}
	\label{MethodsTargetTrackingTE}
	Consider an agent moving with speed $v$ attempting to find a target. We define the tracking error, $\psi$, as the angular difference between the agent's current heading $\phi$ and the allocentric bearing to the target $T_{target}$: $\psi = \phi - T_{target}$. Therefore, $\psi = 0$ corresponds to a direct, straight-line approach.
	The dynamics of $\psi$ are governed by two competing torques.
	Internal neuro-sensory torque ($\tau_{internal}$) is the ``steering wheel" of the agent, derived from the gradient of the Harmonic Hamiltonian. Suppressing the noise term, the internal torque reads as follows (\red{S.4}):
	$$\tau_{internal} = -\eta \frac{\partial H_{\mathrm{eff}}}{\partial \phi} = -\eta \sum_{n=1}^{\infty} n K_n \sin(n\psi).$$
	The kinematic torque, $\tau_{kinematic}$, is a geometric consequence of motion. As the agent moves, the bearing to the target changes (parallax). To maintain a constant relative angle to the target, the agent must turn. The contribution of changing bearing to the tracking error depends on the distance $r$: $\tau_{kinematic} = \frac{v}{r} \sin(\psi)$ (\red{S.4}).
	The full equation of motion for the tracking error is:

	$$\frac{d\psi}{dt} = \tau_{kinematic} + \tau_{internal} = \frac{v}{r} \sin(\psi) - \eta \sum_{n=1}^{\infty} n K_n \sin(n\psi).$$

	The angular nullclines satisfy $\dot{\psi}=0$ at fixed distance $r$, while $\dot r=-v\cos\psi$. For the $n$th harmonic, the angular equation has at most $2n$ isolated simple roots with alternating stability; these do not by themselves define $n$ stable trajectories (\red{S.4}).

	\subsubsection{Harmonic Theory of Bifurcations in target-seeking} 
	\label{MethodsTargetTrackingBifurcation}
	
	Numerical solutions reveal distinct trajectory types in tracking an attractive stimulus. For large $\sigma$, the agent often exhibits back-and-forth motion towards and away from the target. On the other hand, for small $\sigma$, the agent exhibits spiral motion towards the target, followed by a direct escape when it becomes too close to the target. To understand bifurcations between these trajectory types, we analyze the stability of the direct approach and direct escape (see \red{S.4} for details).
	
	The trivial fixed point $\psi = 0$ corresponds to the direct approach. Its stability can be determined by requiring the Jacobian $J = \frac{d\dot{\psi}}{d\psi}\Bigg|_{\psi=0} = \frac{v}{r} - \eta \sum_{n=1}^{\infty} n^2 K_n$, to be smaller than $0$. This leads to the stability condition: $\frac{v}{r} < \eta \sum_{n=1}^{\infty} n^2 K_n$. This stability condition defines a bifurcation radius $r_c$ where the forces balance: $r_c = \frac{v}{\eta \sum n^2 K_n}$. $r_c$ is positive for finite $\sigma$ and $0<W<2\pi$, and diverges as $\sigma\to0$ or $\sigma\to\infty$ (\red{S.4}). Consequently, for small $\sigma$, the neural torque can become too small to correct small errors, and the agent does not exhibit a direct approach. Curved approaches need not maintain a constant tracking error (\red{S.4}).
	
	Similarly, we can determine the stability of the direct escape, $\psi^*=\pi$. For direct escape, we have $J = \frac{d\dot{\psi}}{d\psi}\Bigg|_{\psi=pi} = -\frac{v}{r} - \eta \sum_{n=1}^{\infty} n^2 K_n \cos(n\pi)$.
	The stability condition for the direct escape is then: $\frac{v}{r} >- \eta \sum_{n=1}^{\infty} n^2 K_n \cos(n\pi)$. Therefore, direct escape is stable when $r$ is smaller than $r_e= - \frac{v}{\eta \sum_{n=1}^{\infty} n^2 K_n \cos(n\pi)}$.
	
	$r_e$ is positive for finite $\sigma$ and $0<W<2\pi$ (\red{S.4}), leading to instability of direct escape for large distances. This leads to turning back towards the target at a positive critical distance, $r_e$. This can explain the back-and-forth motion observed for large $\sigma$, as resulting from spatial, distance-dependent bifurcations, through which the direct approach and direct escape alternatively lose and gain stability (\red{S.4}). $r_e$ diverges as $\sigma\to0$ or $\sigma\to\infty$, indicating that for small $\sigma$, direct escape can remain stable even far away from the target, underlying poor navigation for small $\sigma$ (\red{S.4}).
	
	These bifurcations can also be understood in terms of the battle of harmonics. From the stability condition for the direct approach, it is clear that harmonics with $K_n > 0$ contribute to the potential well that traps the agent in a direct approach. On the other hand, harmonics with $K_n < 0$ create a potential hill at $\psi=0$, pushing the agent away. The sign of $K_n$ depends on the decision filter $M_n$, which in turn depends on the decision uncertainty, $W$. While for the first harmonic $M_1$ remains always positive, and thus this harmonic is the only mode whose sign can be flipped only by the ecological interaction kernel, $J_{\mathrm{int}}$, distinguishing attractive from repulsive stimuli, other harmonics can change sign depending on the decision uncertainty ($W$) (\red{S.4}). Because the first harmonic is the main force behind target-seeking, a large value of $\sigma$, by attenuating the contribution of higher harmonics, facilitates direct approach, while a smaller $\sigma$ leads to more complex spiral trajectories due to the contribution of higher harmonics.
	
	\subsection{Harmonic Theory of Bifurcations in binary choice}
	\label{MethodsBifurcations}
	
	In this Section, we develop a mathematical theory of bifurcations in the binary choice problem, where the agent faces two equal targets.
	\subsubsection{The effective potential landscape}
	
	We model the two targets as generating a combined sensory field. The effective potential $U(\phi; y)$ describing the agent's heading at a distance $y$ from a baseline is derived by summing the harmonic contributions from both targets. Using trigonometric sum-to-product identities, the potential simplifies to (see \red{S.6.1}):
	
	$$U(\phi; y) = -2 \sum_{n=1}^{\infty} K_n \cos\left( n \frac{\Delta(y)}{2} \right) \cos\left( n (\phi - \frac{\pi}{2}) \right)$$
	Here, $\Delta(y)$ is the angular separation between the targets as seen from the agent's position. This equation reveals that the potential is a Fourier series in the deviation from the symmetric heading, $(\phi - \pi/2)$. The agent's progress, parameterized by $y$, does not shift the center of the potential but rather modulates the amplitudes of its Fourier components via the term $\cos(n \Delta(y) / 2)$. The decision-making process is thus transformed into a problem of analyzing the stability of a potential whose shape is parametrically driven by the agent's position.
	
	\subsubsection{The critical transition and the nature of the bifurcation}
	
	The transition from averaging to choosing is a symmetry-breaking bifurcation. It occurs when the curvature of the total potential at the center ($\phi = \pi/2$) flips from positive (stable valley) to negative (unstable hill). This is defined by the condition (\red{S.6.4}):
	
	\begin{align}
		\sum_{n=1}^{\infty} n^2 K_n \cos\left( n \frac{\Delta(y_c)}{2} \right) = 0
		\label{eq:eqbifurcationstability}
	\end{align}
	This equation defines the critical angular separation $\Delta(y_c)$ at which the frozen-position symmetric heading loses stability. The phase diagram presented in Fig. \ref{Fig2}\textbf{D} results from a real-space evaluation of this equation, with numerical resolution limits described in \red{S.6.4}. The loss of stability of the symmetric trajectory occurs where the left-hand side of Eq. \eqref{eq:eqbifurcationstability} becomes negative. For the first few harmonics (small $n$), the trajectory-dependent term, $\cos\left( n \frac{\Delta(y_c)}{2} \right)$, is positive far from the target. Thus, as long as the neurosensory term, $K_n$, is positive, all these harmonics are initially stabilizing forces. In $\red{S.6.4}$, we outline the sign of this term as a function of $W$. Namely, $K_n$ is always positive for $n=1$, and it is positive for $n=2$ provided $0<W<\pi$. Thus, both these harmonics are stabilizing. When the agent moves closer to the target, $\cos\left( n \frac{\Delta(y_c)}{2} \right)$ can become negative for $n\geq2$, making these harmonics destabilizing forces. As the agent moves in space, the sign of the trajectory-dependent term alternates, causing the alternation of each harmonic between stabilizing and destabilizing forces. The speed of alternation increases with $n$. The full sum, however, must be evaluated before inferring additional bifurcations. 
	
	The subcritical or supercritical nature of an ordinary pitchfork bifurcation can be read out based on the sign of the fourth derivative at the bifurcation point (\red{S.6.4}):
	\begin{equation}
		\left. \frac{\partial^4 U}{\partial \phi^4} \right|_{\phi=\pi/2} = -2 \sum_{n=1}^{\infty} n^4 K_n \cos\left(n\frac{\Delta(y)}{2}\right)
		\label{eq:fourth_deriv}
	\end{equation}
	
	A positive value gives a supercritical bifurcation, whereas a negative value gives a subcritical bifurcation \cite{Strogatz2024}; a zero value requires higher-order analysis. The numerically resolved cases in Fig. \ref{Fig2}\textbf{D(ii)} are supercritical (\red{S.6.4}).
	
	The dependence of the bifurcation point on the bump width can also be understood in terms of the battle of harmonics. A narrow bump delays the first zero of $\sin(nW/2)$ (contained in $K_n$). This gives significant weight to the higher harmonics ($n \geq 2$), which are the primary drivers of instability. A system that strongly represents these splitting harmonics will satisfy the bifurcation condition at a smaller angular separation $\Delta$, leading to an earlier decision.

	\subsubsection{Turing-like pattern formation in heading space}
	
	Harmonic formulation suggests that spatial decision-making is analogous to Turing pattern formation in reaction-diffusion systems \cite{Turing1990}, but occurring in the space of heading direction. The harmonics act like ``morphogens". The fundamental harmonic ($n=1$) creates a single potential well centered at the average direction ($\phi = \pi/2$). It promotes consensus and averaging. However, higher harmonics can change sign as the agent moves in space, alternating between stabilizing (activators) and destabilizing (inhibitors) forces (\red{S.6.3}). 
	
	While similarities with Turing pattern formation are manifest in the competition between stabilizing and destabilizing harmonics, differences also exist. In the Harmonic Pattern Formation, the stabilizing nature of harmonics is a function of the geometrical representation of space in the brain. As the agent moves in space, harmonics, $n\geq2$, alternate between stabilizing and destabilizing forces. This leads to a richer pattern formation for smaller $\sigma$, where higher harmonics contribute more.

	\subsection{The parity principle and the harmonic modes of collective behavior}
	\label{MethodsParityPrinciple}
	
	In this Section, we provide a concise mathematical classification of harmonic building blocks of collective motion patterns. This analysis is detailed and numerically confirmed in \red{S.9}.
	\subsubsection{Odd harmonics: conflict and frustration}
	
	To classify the behavior of harmonics, we assume $M_n > 0$, so that $K_n > 0$ for attractive interactions and negative for repulsive interactions. This assumption is always valid for the first and second harmonics (assuming $0<W<\pi$), but not necessarily for higher harmonics (resulting in the fact that higher harmonics cannot distinguish repulsive and attractive interactions). However, for our classification system, we make this choice as a convention.
	
	For attractive interactions, minimizing Eq. \eqref{eq:odd}, we have (\red{S.9}):
	
	$$
	\cos(n(\phi^{(a)} - T^{(a,b)})) \to +1 \quad \Rightarrow \quad n(\phi^{(a)} - T^{(a,b)}) \approx 2k\pi \quad \Rightarrow \quad \phi^{(a)}_k \approx T^{(a,b)} + \frac{2k\pi}{n},
	$$
	$$
	\cos(n(\phi^{(b)} - T^{(a,b)})) \to -1 \quad \Rightarrow \quad n(\phi^{(b)} - T^{(a,b)}) \approx (2m+1)\pi \quad \Rightarrow \quad \phi^{(b)}_m \approx T^{(a,b)} + \frac{(2m+1)\pi}{n},
	$$ 
	
	where $k,m\in\{0,..,n-1\}$. For repulsive interactions, ($K_n<0$), minimizing Eq. \eqref{eq:odd}, we have:
	
	$$
	\cos(n(\phi^{(a)} - T^{(a,b)})) \to -1 \quad \Rightarrow \quad n(\phi^{(a)} - T^{(a,b)}) \approx (2k+1)\pi \quad \Rightarrow \quad \phi^{(a)}_k \approx T^{(a,b)} + \frac{(2k+1)\pi}{n}
	$$
	$$
	\cos(n(\phi^{(b)} - T^{(a,b)})) \to +1 \quad \Rightarrow \quad n(\phi^{(b)} - T^{(a,b)}) \approx (2m)\pi \quad \Rightarrow \quad \phi^{(b)}_m \approx T^{(a,b)} + \frac{(2m)\pi}{n}
	$$
	As can be seen, in both cases, consensus regarding heading directions is not possible, and the agents should maintain an $n$-dependent phase difference with respect to their common bearing. In larger groups, simultaneous minimization of all pair-wise potentials depends on the bearing geometry, for either parity. Geometric incompatibilities can contribute to fission-fusion and frustrated dynamics (\red{S.9}).
	
	\subsubsection{Even harmonics: order and consensus}
	For attractive interactions, minimizing Eq. \eqref{eq:even} for even harmonics, we have (\red{S.9}):
	$$
	\cos(n(\phi^{(a)} - T^{(a,b)})) \to +1 \quad \Rightarrow \quad n(\phi^{(a)} - T^{(a,b)}) \approx 2k\pi \quad \Rightarrow \quad \phi^{(a)}_k \approx T^{(a,b)} + \frac{2k\pi}{n},
	$$
	$$
	\cos(n(\phi^{(b)} - T^{(a,b)})) \to +1 \quad \Rightarrow \quad n(\phi^{(b)} - T^{(a,b)}) \approx 2m\pi \quad \Rightarrow \quad \phi^{(b)}_m \approx T^{(a,b)} + \frac{2m\pi}{n}.
	$$
	For the second harmonic, for instance, this leads to alignment (\{$\phi^{(a)}_1=T^{(a,b)}$, $\phi^{(b)}_2=T^{(a,b)}$\} and \{$\phi^{(a)}_2=T^{(a,b)}+\pi$, $\phi^{(b)}_1=T^{(a,b)}+\pi$\}), and anti-alignment (\{$\phi^{(a)}_2=T^{(a,b)}+\pi$, $\phi^{(b)}_2=T^{(a,b)}$\} and \{$\phi^{(a)}_1=T^{(a,b)}$, $\phi^{(b)}_1=T^{(a,b)}+\pi$\}). 
	
	For repulsive interactions:
	$$
	\cos(n(\phi^{(a)} - T^{(a,b)})) \to -1 \quad \Rightarrow \quad n(\phi^{(a)} - T^{(a,b)}) \approx (2k+1)\pi \quad \Rightarrow \quad \phi^{(a)}_k \approx T^{(a,b)} + \frac{(2k+1)\pi}{n},
	$$
	$$
	\cos(n(\phi^{(b)} - T^{(a,b)})) \to -1 \quad \Rightarrow \quad n(\phi^{(b)} - T^{(a,b)}) \approx (2m+1)\pi \quad \Rightarrow \quad \phi^{(b)}_m \approx T^{(a,b)} + \frac{(2m+1)\pi}{n}.
	$$
	For the second harmonic for instance, this leads to perpendicular alignment (\{$\phi^{(a)}_1=T^{(a,b)}+\pi/2$, $\phi^{(b)}_2=T^{(a,b)}+\pi/2$\} and \{$\phi^{(a)}_2=T^{(a,b)}+3\pi/2$, $\phi^{(b)}_1=T^{(a,b)}+3\pi/2$\}) and opposed transverse preferences associated with milling (\{$\phi^{(a)}_1=T^{(a,b)}+\pi/2$, $\phi^{(b)}_1=T^{(a,b)}+3\pi/2$\} and \{$\phi^{(a)}_2=T^{(a,b)}+3\pi/2$, $\phi^{(b)}_2=T^{(a,b)}+\pi/2$\}).

	\subsection{Numerical methods}
	
	For figures based on direct evaluation of the theory (in Figs. \ref{Fig0}\textbf{B}-\textbf{D}, \ref{Fig2}\textbf{B}-\textbf{C}, \ref{Fig3}\textbf{B(i)}, \ref{Fig3}\textbf{C(i)}, \ref{Fig3}\textbf{D}, \ref{Fig4}\textbf{A(i)}-\textbf{D(i)}, and \ref{Fig4}\textbf{A(iv)}-\textbf{D(iv)}) the displayed quantities were computed from the closed-form harmonic couplings $
	K_n = J_{\mathrm{int}}\, c_n(\sigma)\, M_n(W)$, 
	$c_n(\sigma)=\frac{1}{\pi}\exp\!\left(-\frac{n^2\sigma^2}{2}\right)$, and
	$M_n(W)=\frac{4}{n\pi}\sin\!\left(\frac{nW}{2}\right)$,
	with harmonic sums truncated at the value of \(n_{\max}\) stated in each figure caption. Potential landscapes were generated by direct numerical evaluation of the corresponding effective Hamiltonian on a dense angular grid. For single-agent potential plots in the binary-choice and binary-escape problems (Figs. \ref{Fig2}\textbf{C}, \ref{Fig3}\textbf{B(i)}, and \ref{Fig3}\textbf{C(i)}) and the bifurcation diagrams (Figs. \ref{Fig2}\textbf{B} and \ref{Fig3}\textbf{D}), the angular coordinate \(\phi\) was discretized uniformly, and the potential was evaluated at each grid point using the exact analytical expressions given in Eq. \eqref{eq:eqHarmonicPotentialBifurcation}. 
	
	The bifurcation maps for the binary-choice problem in Fig. \ref{Fig2}\textbf{D} were generated by numerical evaluation of the stability of the central (symmetric) branch. To avoid cancellation, derivatives were evaluated with signed-logarithmic wrapped-Gaussian image sums for narrow kernels and convergent Fourier sums for broader kernels (\red{S.6.4}). The parameter plane \((W,\sigma)\) was sampled on an \(80\times 120\) grid with \(\sigma \in [0.01,1.0]\) and \(W \in [0.01,\pi+0.01]\). For each parameter pair, the central-branch stability function \(C(\Delta)\) was evaluated by scanning the interval \(\Delta \in [0,\pi]\). The local bifurcation type was classified from the fourth derivative at numerically resolved critical points. Roots too close to $\Delta=W$ to resolve separately were left unclassified (\red{S.6.4}).
	
	Trajectory-density panels in the single-agent figures were constructed from repeated simulations with initial headings reported in the figures. Dependence on initial headings can be strong for small $\sigma$ (\red{S.6.4}).
	
	Both individual and collective motion simulations were performed by numerical solutions of the effective Hamiltonian dynamics. For each agent \(i\), the total torque was computed by summing pairwise contributions from all other agents (if applicable) and all targets (if applicable), with each contribution decomposed into harmonics up to the specified \(n_{\max}\). At each time step, the heading was updated according to
	\(
	\phi^{(a)}(t+\Delta t)=\phi^{(a)}(t)+\eta\,\tau^{(a)}\,\Delta t+\sqrt{2D_r\Delta t}\,\xi_a,
	\)
	where \(\tau^{(a)}=-\frac{\partial H_{\mathrm{eff}}^{(a)}}{\partial \phi^{(a)}}\) is the total deterministic torque and \(\xi_a\) is a Gaussian random variable of zero mean and unit variance. Positions were then updated with constant speed \(v_0^{\mathrm{eff}}\) using the new heading. Periodic boundary conditions were used when indicated. Unless otherwise stated, we set $\Delta t=0.1$. In collective motion simulations, initial positions were sampled uniformly in the domain, and initial headings were chosen uniformly at random in \([0,2\pi]\).

	\subsection{Measure of collective motion}
	
	To investigate collective motion patterns, we have used several established measures of order and structure in collective motion. These include global and local angular and vectorial order parameters, global and local nematic order, mean nearest-neighbor distance, all-pairs distance, and normalized angular momentum. Extensive analysis using these metrics is presented in the \red{Supplementary Information}. In the main text, we have presented results for a limited set of these measures. As a measure of global order, we use the global angular order parameter (AOP). This is defined as the magnitude of the mean unit-heading vector, $\mathrm{AOP}(t)=\left|\frac{1}{N}\sum_{i=1}^{N}\hat{\mathbf{u}}_i(t)\right|$, where the unit heading vector is $\hat{\mathbf{u}}_i(t)=\frac{\mathbf{v}_i(t)}{|\mathbf{v}_i(t)|}$. Here, $\mathbf{v}_i(t)$ is the agent $i$’s velocity. This metric approaches $1$ when all agents move in the same direction and approaches $0$ when headings are disordered.
	
	While global order distinguishes between parallel and antiparallel motion, and thus takes its maximum when all the agents are aligned in the same direction, nematic order does not. Consequently, it is maximized when agents' movement axis, but not necessarily their heading directions, are aligned. This quantity is defined as $S_{\mathrm{global}}(t)=\sqrt{\left\langle u_x^2-u_y^2 \right\rangle^2+\left\langle 2u_xu_y\right\rangle^2}$. The mathematical derivation of this formula is provided in \red{S.13}.
	
	Finally, to detect rotational states, we computed the normalized angular momentum about the instantaneous center of mass, $\Lambda(t)=
	\frac{|M(t)|}
	{\sum_{i=1}^{N}
		\|\mathbf{r}_i(t)-\mathbf{r}_{\mathrm{cm}}(t)\|\,\|\mathbf{v}_i(t)\|}$, where the angular momentum is defined as, $
	M(t)=\sum_{i=1}^{N}
	\left[
	\left(\mathbf{r}_i(t)-\mathbf{r}_{\mathrm{cm}}(t)\right)\times \mathbf{v}_i(t)
	\right]_z$. Here $\mathbf{r}_i(t)$ is the agent $i$'s position vector and $\mathbf{r}_{\mathrm{cm}}$ is the population's center of mass coordinate vector. See \red{S.13} for details and a more extensive set of measures.

	\section{Acknowledgements}
	We acknowledge the use of AI tools for programming assistance, proofreading the manuscript, and analytical consistency checks. The authors take full responsibility for the scientific content, analyses, interpretations, and conclusions.  
	
	The authors acknowledge funding from Deutsche Forschungsgemeinschaft (DFG, German Research Foundation) under Germany’s Excellence Strategy - EXC 2117-422037984, the Deutsche Forschungsgemeinschaft Gottfried Wilhelm Leibniz Prize 2022 584/22 (I.D.C.), the Max Planck Society, DFG project number 462886202, the European Union’s Horizon 2020 Research and Innovation Programme under the Marie Skłodowska-Curie Grant agreement no. 860949, the Struktur- und Innovations fonds für die Forschung of the State of Baden-Württemberg, the Pathfinder European Innovation Council Work Programme no. 101098722, the Office of Naval Research Grant N0001419-1-2556.

	\pagebreak
	\clearpage
	\newpage
	
	\begin{figure}
		\centering
		\includegraphics[width=1\linewidth, trim = 10 50 10 10, clip,]{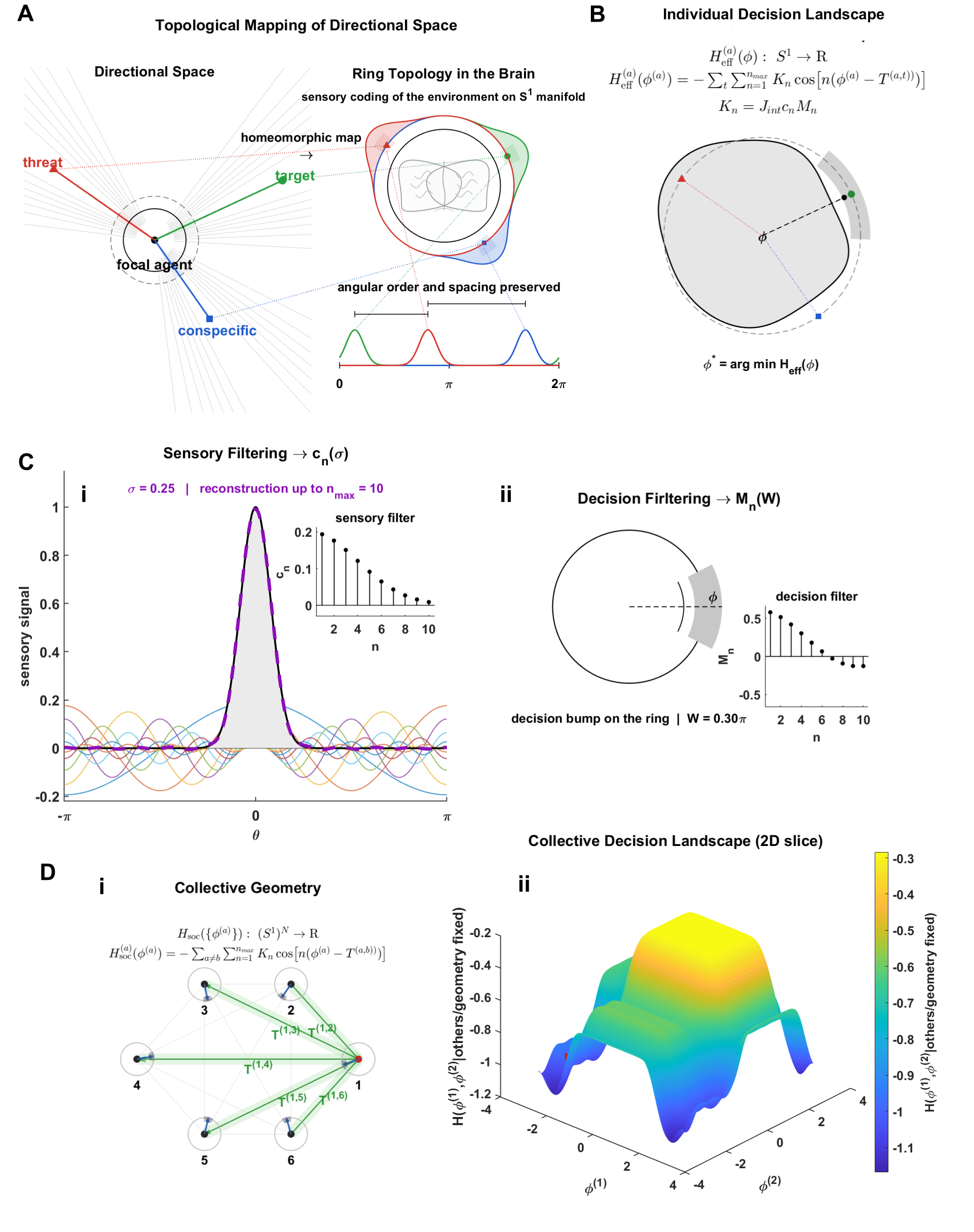}
	\end{figure}

	\begin{figure}
		\centering
		\caption{From topological representation of directional space to harmonic decision landscapes in individuals and collectives. 
			\textbf{(A)}: The topological ring provides a homeomorphic mapping of directional space. Each environmental cue is represented by a localized Gaussian bump on the ring manifold \(S^1\) such that angular ordering and relative spacing are preserved by construction. 
			\textbf{(B)}: The theory is built from four assumptions: (1) sensory inputs are integrated additively; (2) spatial information is represented on a topological mapping of space, taken here to be a ring manifold \(S^1\) for directional decisions in two dimensions; (3) the current decision state is encoded by a localized activity bump of width \(W\) on that manifold; and (4) neural bump-shape dynamics are fast relative to overt movement, allowing microscopic neural variables to be coarse-grained. Under these assumptions, the neural dynamics reduce to a \textit{decision landscape}: an effective Hamiltonian \(H_{\mathrm{eff}}^{(a)}(\phi):S^1\rightarrow \mathrm{R}\) over heading space, whose minimum \(\phi^*=\arg\min H_{\mathrm{eff}}(\phi)\) gives an instantaneous preferred heading. The decision landscape is a superposition of harmonic modes of the angular deviation of the agent's heading and bearing to environmental stimuli, where each mode $n$ is weighted by a neuro-ecological factor, \(K_n = J_{\mathrm{int}}\, c_n\, M_n\). This factor is a product of two filters, a sensory filter, $c_n$, and a decision filter, $M_n$, and a (possibly distance-dependent) constant, $J_{\mathrm{int}}$, representing the attractive or aversive nature of the environmental stimuli.
			\textbf{(C)} The sensory filter arises from the spectral decomposition of the sensory input. In \textbf{(i)} a sensory Gaussian profile on the ring is decomposed into harmonic modes, whose amplitudes define the sensory filter $c_n(\sigma)=\frac{1}{\pi}\exp\big(-\frac{n^2\sigma^2}{2}\big)$. The overlaid colored curves indicate the first $10$ harmonics, and their superposition (up to $n_{\mathrm{max}}=10$) reconstructs the sensory bump. The decision filter, $M_n(W)=\frac{4}{n\pi}\sin\big(\frac{nW}{2}\big)$, arises from the spectral decomposition of the activity bump with width (uncertainty) $W$, representing the decision state (\textbf{(ii)}). This filter selects and reweights harmonic contributions based on the agent's decision representation on the ring manifold.
			\textbf{(D)}: The theory is straightforwardly extended to collectives by considering a collective of agents perceiving each other as social stimuli (\textbf{i}). The collective decision landscape (Hamiltonian) is built from pairwise interactions through the bearings \(T^{(a,b)}\), yielding a high-dimensional decision landscape which lives on the product space \((S^1)^N\). A two-dimensional slice of this landscape is shown (\textbf{ii}), together with the corresponding six-agent geometry (\textbf{i}). Green edges highlight the bearings from agent \(1\) to the rest of the group. All illustrations (in \textbf{B} and \textbf{D}) result from numerical solutions of the Harmonic Hamiltonian using parameter values: $\sigma=0.25$, $W=0.3\pi$, truncating the harmonic sums at $n_{\mathrm{max}}=10$, and setting $J_{\mathrm{int}}=1$. The same parameter values are used for illustrating sensory filtering and decision filtering in \textbf{C}.}
		\label{Fig0}
	\end{figure}

	\begin{figure}
		\centering
		\includegraphics[width=1\linewidth]{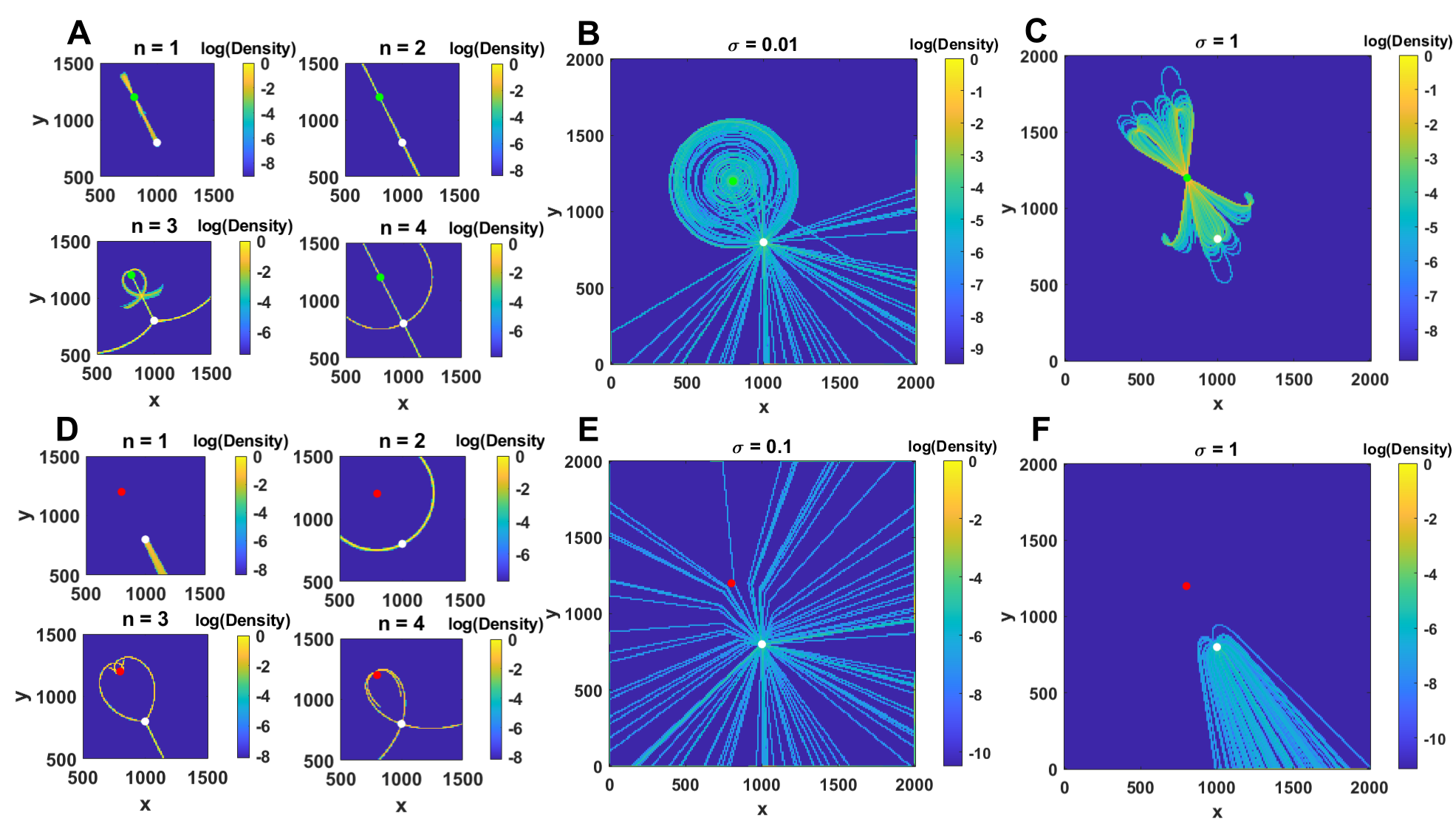}
		\caption{Harmonic Theory of target-seeking and avoidance. \textbf{A}: The density map of trajectories of an agent starting from the white circle and facing an attractive stimulus (green circle) for the first four harmonics is plotted. The $n$th harmonic has $n$ fixed-bearing minima at $\psi=2k\pi/n$, $k\in\{1,\ldots,n\}$, with respect to bearing to the target. This leads to a direct approach for the first harmonic and alignment (where both direct approach, $\psi=0$, and direct escape, $\psi=\pi$, are possible) for the second harmonic. \textbf{B} and \textbf{C}: The agent's trajectories result from the superposition of harmonics controlled by neuro-sensory parameters, $W$ and $\sigma$. For small $\sigma$, the contribution of higher harmonics is weighted more strongly, leading to a spiral approach to the target (\textbf{B}). The attenuation of higher harmonics for large $\sigma$ leads to a direct approach to the target and back-and-forth motion resulting from bifurcations between direct escape and direct approach, once the agent reaches the target (\textbf{C}). \textbf{D}: The density map of trajectories of the agent when facing a repulsive stimulus (red circle) for the first four harmonics is plotted. For repulsive interactions, the $n$th harmonic has $n$ fixed-bearing minima at $\psi=(2k+1)\pi/n$, $k\in\{1,\ldots,n\}$, with respect to bearing to the target. This leads to direct escape for the first harmonic and perpendicular preferred headings ($\psi=\pi/2$ or $3\pi/2$ at fixed bearing, which do not imply circular motion) for the second harmonic. \textbf{E} and \textbf{F}: For small $\sigma$, high contribution of higher harmonics degrades target-avoidance, induced by the first harmonic, leading to late turning away from the target (\textbf{E}). The attenuation of higher harmonics for large $\sigma$ leads to direct escape from the target early on (\textbf{F}). Parameter values: $v_0^{\mathrm{eff}}=1$ and $\eta=0.1$. In \textbf{A} to \textbf{C}, $h=1$, and in \textbf{D} to \textbf{F}, $h=-1$. In \textbf{A} and \textbf{D}, $\sigma=0.1$ and $W=0.903$, and in other panels, $W=3\pi/5$ and $n_{\mathrm{max}}=1024$. The deterministic dynamics ($D_r=0$) is considered.}
		\label{Fig1}
	\end{figure}
	
	\begin{figure}
		\centering
		\includegraphics[width=1\linewidth]{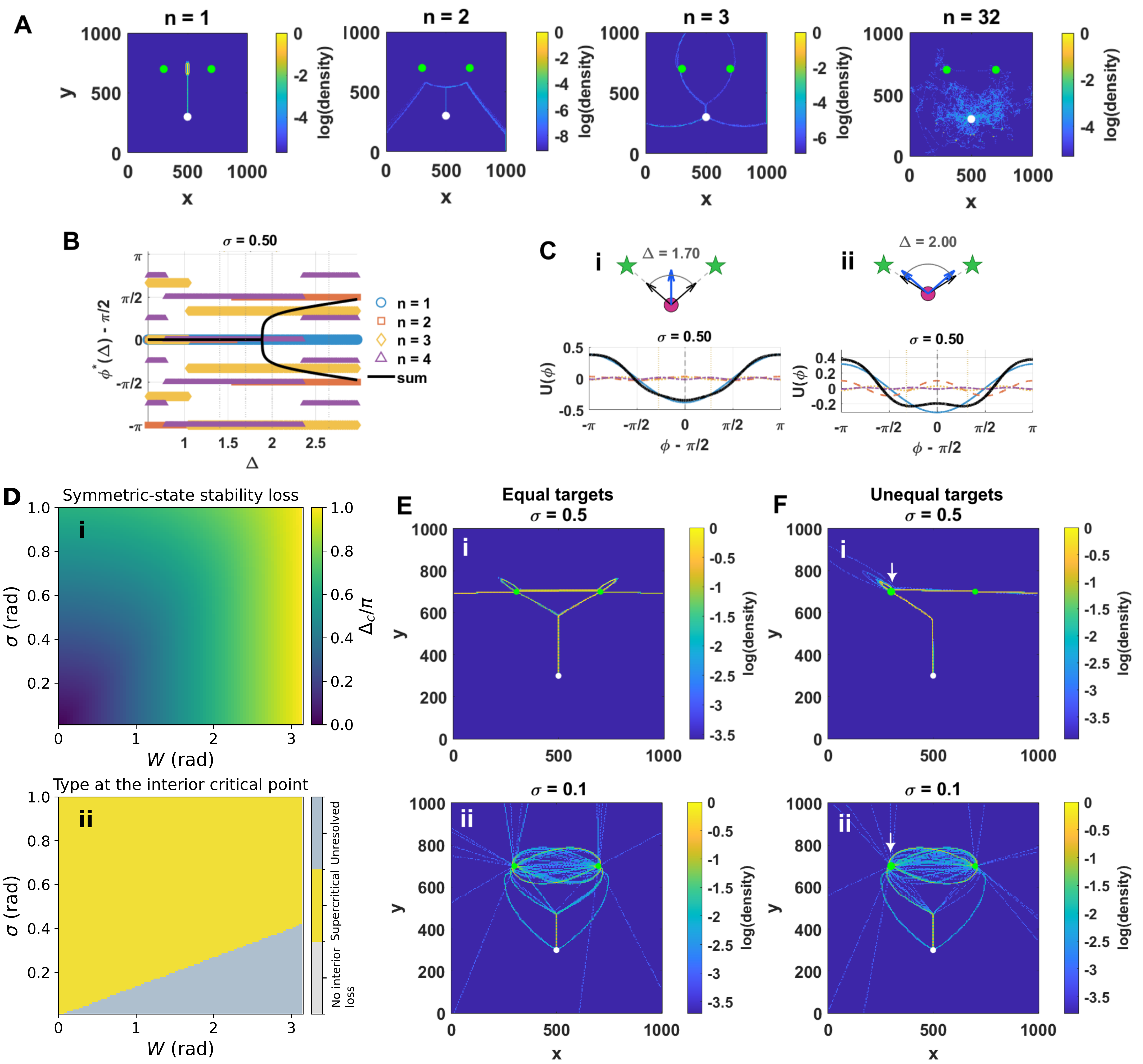}
	\end{figure}

	\begin{figure}
	\centering
	\caption{Binary Choice Problem in the Harmonic Theory. \textbf{A}: The trajectories resulting from individual harmonics are presented. For the first harmonic, averaging remains stable, leading to a movement towards and staying close to the average target. For higher harmonics, averaging can change stability as the agent moves, leading to bifurcations. The post-bifurcation trajectory depends on the harmonic, with more possible trajectories observed for higher harmonics. This multiplicity of solutions for too high harmonics can induce a random-walk-like behavior. \textbf{B}: The stable heading direction along the symmetry axis, parametrized by $\Delta(y)$, is plotted for the first four harmonics and the sum of harmonics up to $n_{\mathrm{max}}=512$ for a large $\sigma$. For each harmonic $n>1$, a bifurcation occurs as $\Delta$ increases. The superposition of harmonics gives rise to a supercritical bifurcation, where two stable solutions corresponding to a choice appear. \textbf{C}: The potential landscape for two angular separations, $\Delta$, chosen below and above the bifurcation point, is shown. The first four harmonic potentials and the potential sum up to $n_{\mathrm{max}}=512$ are shown. While individual harmonics exhibit several minima and maxima, with the number of minima increasing with harmonic number, for large $\sigma$, the potential landscape simplifies and exhibits only one minimum for small angular separation, $\Delta$, (far from the targets), corresponding to a symmetric trajectory. As the agent approaches the target ($\Delta$ increases), this minimum loses stability, and two minima, each inclined towards one of the targets, appear continuously in a supercritical bifurcation. \textbf{D}: The first (smallest interior value) critical angular separation where bifurcation occurs (\textbf{i}), and its bifurcation type (\textbf{ii}) are presented as a function of the bump width, $W$, and sensory kernel, $\sigma$. The bifurcation angle increases with $W$, indicating that higher decision uncertainty leads to a later decision. Numerically resolved bifurcations of the frozen square-bump, Gaussian sensory model are supercritical. Unresolved cases are marked separately; gray cells have no interior loss of stability (\red{S.6.4}). \textbf{E}: Examples of decision-making patterns for large $\sigma$ (\textbf{i}) and small $\sigma$ (\textbf{ii}). For small $\sigma$, multiple trajectories and rich pattern formation are observed (\red{S.6.4}). \textbf{F}: When the agent faces two unequal targets, the attenuation of higher harmonics for large $\sigma$ leads to accurate decisions. However, for small $\sigma$, the agent makes faster, but less accurate decisions. Parameter values: In \textbf{A}: $v_0^{\mathrm{eff}}=1$, $h_1=h_2=1$, $W=3\pi/5$, and $\eta=5$. In \textbf{D}: $h_1=h_2=1$. In \textbf{B} and \textbf{C}: $W=3\pi/5$, $\sigma=0.5$, and $n_{\mathrm{max}}=512$ (used in the sum). In \textbf{E}: $v_0^{\mathrm{eff}}=1$, $n_{\mathrm{max}}=512$, $h_1=h_2=1$, $W=3\pi/5$, and $\eta=5$. In \textbf{F}: $v_0^{\mathrm{eff}}=1$, $n_{\mathrm{max}}=128$, $h_1=1$, $h_2=0.99$, $W=3\pi/5$, and $\eta=5$. In \textbf{A}, \textbf{E}, and \textbf{F}, $24$ runs starting from a random initial heading of the agent in the interval $[0,\pi]$ are used. In all panels, the deterministic dynamics ($D_r=0$) are considered.}
	\label{Fig2}
\end{figure}

	\begin{figure}
		\centering
		\includegraphics[width=1\linewidth]{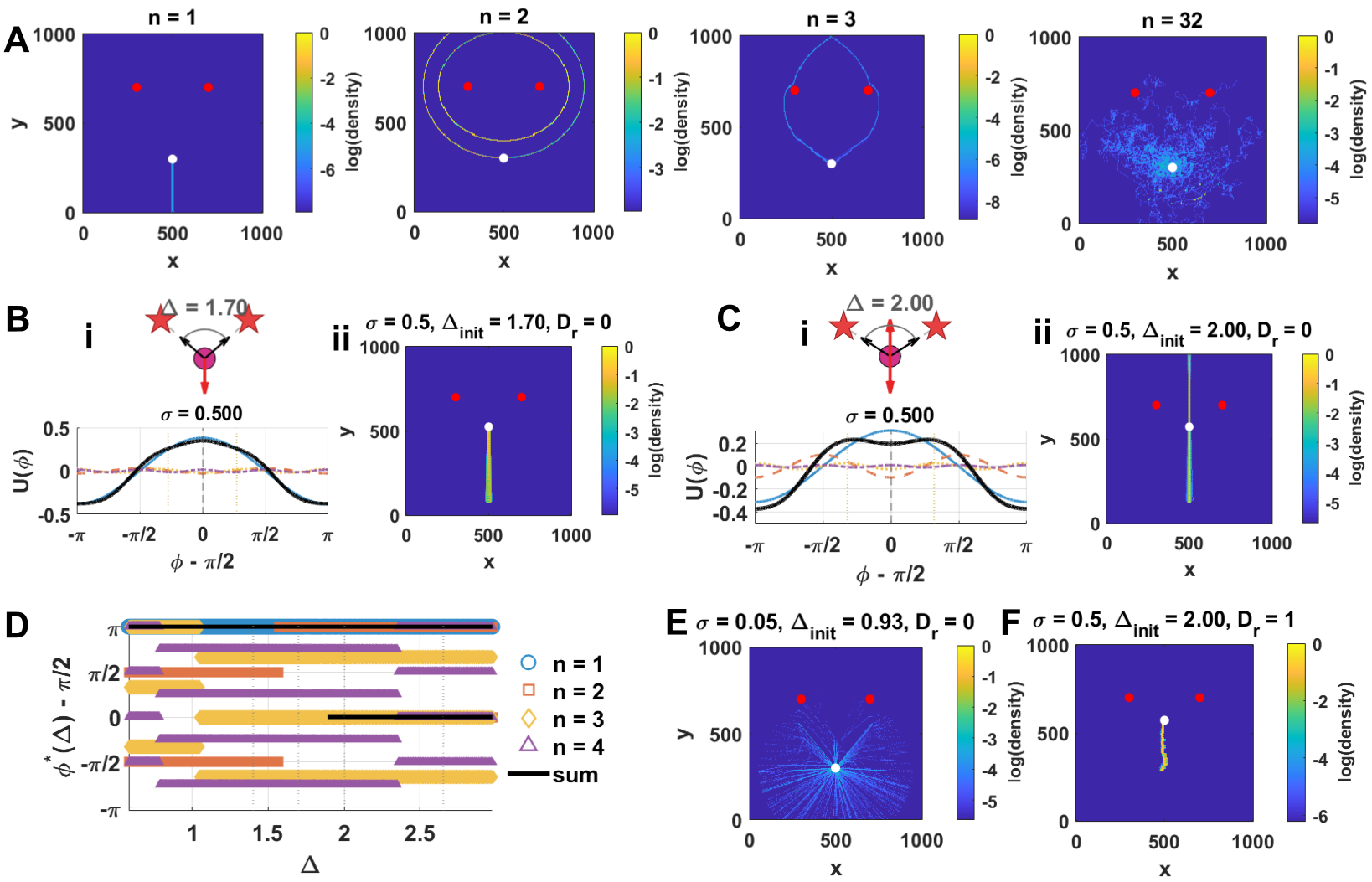}
		\caption{Binary escape problem in the Harmonic Theory. \textbf{A}: The decomposition of behavior into harmonic modes in the binary escape problem, where the agent faces two equal repulsive stimuli, reveals that the first harmonic is the fundamental mode leading to escape from average repulsive stimuli, while higher harmonics often lead to movement towards the repulsive stimuli, or a random-walk-like behavior for very high harmonics. \textbf{B} and \textbf{C}: The potential landscape when the agent is located along the symmetry axis and for two different values of angular separation of the targets, $\Delta$, is presented. The attenuation of higher harmonics for large $\sigma$ leads to a simplified potential landscape. Far away from the repulsive stimuli (small $\Delta$ in \textbf{B}), the agent exhibits direct escape along the symmetry axis, $\phi=3\pi/2$. \textbf{B(i)} shows the potential landscape exhibiting a single minimum at $\phi=3\pi/2$, corresponding to direct escape. In \textbf{B(ii)}, a colormap of $240$ runs when the agent's initial position corresponds to an angular separation, $\Delta_{\mathrm{init}}=1.7$, is presented. Closer to the stimuli, implying larger $\Delta_{\mathrm{init}}$, the extremum at $\phi=\pi/2$, corresponding to direct approach along the symmetry axis, becomes stable (\textbf{C(i)}). Colormap of $240$ runs in \textbf{C(ii)} indicates bistability, where both direct escape and direct approach along the symmetry axis are possible. \textbf{D}: The stable heading directions along the symmetry axis for the first four harmonics, and the superposition of harmonics up to $n_{\mathrm{max}}=512$ is shown as a function of $\Delta$. A subcritical bifurcation, above which the system becomes bistable, is observed (black line). 
			 \textbf{E}: For small $\sigma$, higher harmonics contribute more strongly, and the reported trajectories include motion towards the repulsive stimuli. \textbf{F}: A colormap of $240$ runs in the bistable region (the same parameter values as in \textbf{C(ii)}) in the presence of noise is presented. By facilitating escape from approach states, noise improves avoidance behavior. Parameter values: In \textbf{A}, \textbf{B(ii)}, \textbf{C(ii)}, \textbf{E}, and \textbf{F}: $v_0^{\mathrm{eff}}=1$, $h_1=h_2=-1$, $W=3\pi/5$, $\eta=5$, and (other than \textbf{A}) $n_{\mathrm{max}}=512$. In \textbf{B(i)}, \textbf{C(i)}, and \textbf{D}: $h_1=h_2=-1$, $W=3\pi/5$, $n_{\mathrm{max}}=512$. $D_r$ is $0$ in \textbf{A}, and it is shown in other panels (if applicable). In \textbf{A} snapshots show the results of $24$ runs starting from random initial heading in the interval $[0,\pi]$. In other panels (if applicable), $240$ runs starting from random initial heading in the interval $[0,2\pi]$ are used.}
		\label{Fig3}
	\end{figure}
	
	\begin{figure}
		\centering
		\includegraphics[width=1\linewidth]{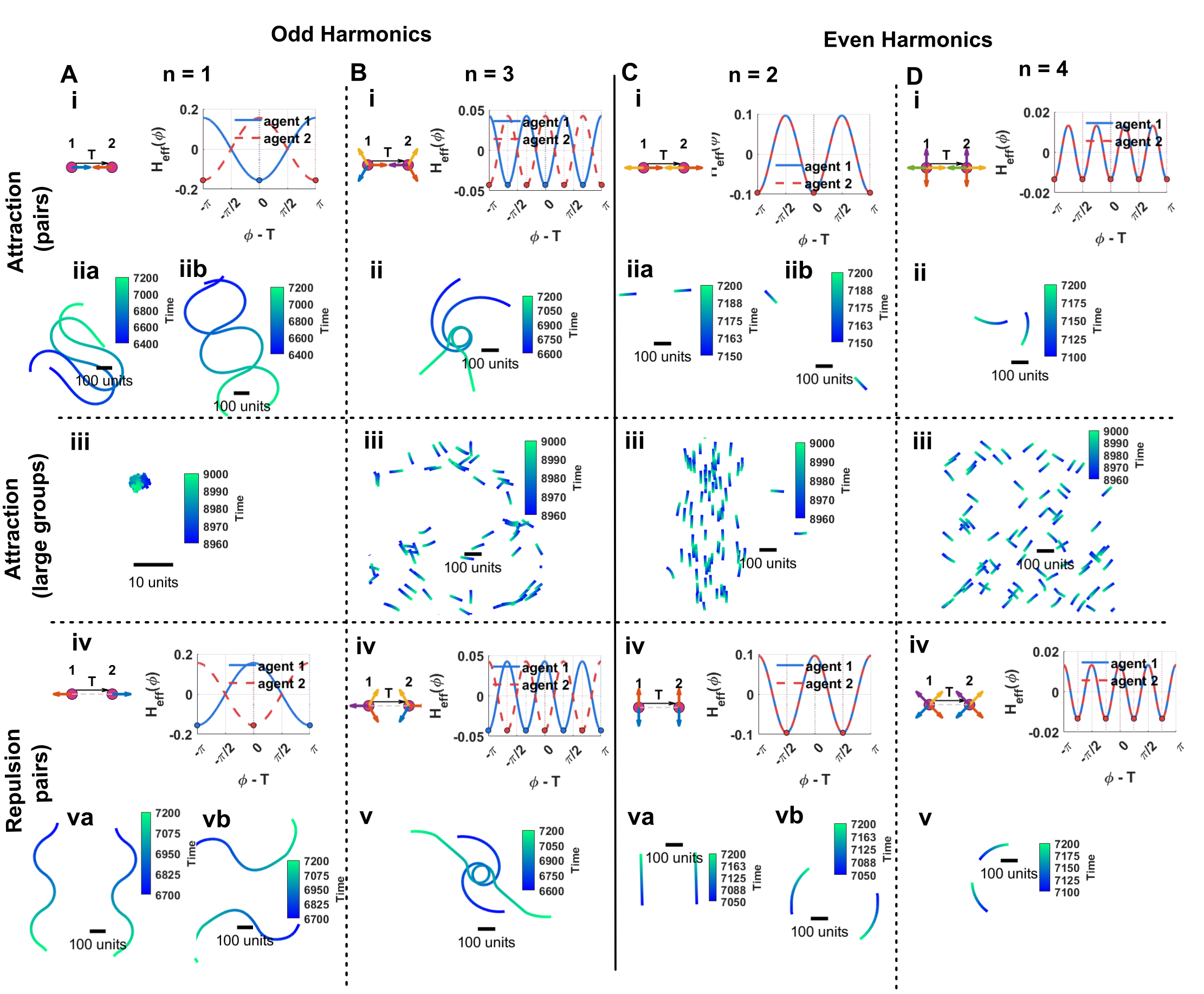}
	%	\caption{}
	\end{figure}
		\begin{figure}
		\centering
		\caption{Collective behavior resulting from harmonics. In \textbf{A(i)} to \textbf{D(i)}, the potential of a pair of agents for attractive interactions as a function of the deviation of their heading direction with respect to the bearing from agent 1 to 2, $\phi-T$, for the first four harmonics, and in a static configuration is plotted. The minimum energy heading direction and the resulting heading vectors are shown. For a harmonic $n$, $n$ possible heading directions exist. For odd harmonics, the potential of the two agents exhibits a phase shift, leading to phase-shifted equilibrium headings of the agents (\textbf{A(i)} and \textbf{B(i)}). By contrast, the bearing symmetry of the even harmonics results in similar potentials and an identical set of heading directions for agents (\textbf{C(i)} and \textbf{D(i)}). To examine dynamic configurations, we present the trajectories of agents, color-coded by time. \textbf{First harmonic}: In pairs, when interactions are attractive, lack of consensus leads to escape-pursuit dynamics (\textbf{A(iia)}) or in-phase synchronized oscillations (\textbf{A(iib)}). In larger groups, agents form frustrated aggregations (\textbf{A(iii)}). Repulsive interactions induce a $\pi/n$ shift in equilibrium headings compared to attractive interactions (\textbf{A(iv)} and \textbf{B(iv)}). In pairs, this leads to synchronization in parallel (\textbf{A(va)}) or perpendicular to the common bearing (\textbf{A(vb)}). \textbf{Third harmonic}: In pairs, lack of consensus drives fission-fusion dynamics with transient rotational milling where the attractive or repulsive nature of interactions only affects the phase of agents with respect to their common bearing, leading to inward (\textbf{B(ii)}) or outward (\textbf{B(v)}) spirals for attractive, and repulsive interactions, respectively. In a large group, the third harmonic leads to fission-fusion dynamics and the formation of vortices (\textbf{B(iii)}). \textbf{Second harmonic}: In pairs, consensus leads to alignment (\textbf{C(iia)}) or anti-alignment (\textbf{C(iib)}). In larger groups, and when interactions are attractive, the second harmonic leads to wide bidirectional columns, exhibiting the reflection symmetry of the second harmonic Hamiltonian (\textbf{C(iii)}). For repulsive interactions, perpendicular alignment, where alignment is perpendicular to the common bearing, is observed. In-phase perpendicular alignment leads to band formation (\textbf{C(va)}) and out-of-phase (i.e., anti-) perpendicular alignment leads to milling (\textbf{C(vb)}). \textbf{Fourth harmonic}: When interactions are attractive, in addition to solutions of the second harmonic, agents can maintain a phase difference $\pi/2$ or $3\pi/2$, leading to configurations such as intermittent inward and outward spirals (\textbf{D(ii)}). In larger groups, the multiplicity of solutions leads to the decomposition of the population into two subgroups ordering among themselves, and often maintaining a phase difference of $\pi/2$ with other subgroups (\textbf{D(iii)}). For repulsive interactions, equilibrium heading shifts by $\pi/4$ (\textbf{D(iv)}). An example of the resulting motion pattern is a milling pattern, associated with preferred headings $\pi/4$ and $7\pi/4$ relative to their common bearing (\textbf{D(v)}). Parameter values: In \textbf{A(i)}-\textbf{D(i)} and \textbf{A(iv)}-\textbf{D(iv)}: $h=(\pm)1$, $W=0.9$, $\sigma=0.5$. In other panels: $v_0^{\mathrm{eff}}=2$, $W=0.9033$, $\sigma=0.1$, $\eta=0.1$, and $D_r=0$. In \textbf{A(ii)}-\textbf{C(ii)}: $h=1$. In \textbf{D(ii)}: $h=10$. In \textbf{A(iii)}-\textbf{D(iii)}: $h=10$. In \textbf{A(v)}-\textbf{B(v)}: $h=-1$. In \textbf{C(v)}-\textbf{D(v)}: $h=-10$. Interactions are distance-independent. Agents move in a periodic space with size $L=1000$.}
		\label{Fig4}
	\end{figure}
	
	\begin{figure}
		\centering
		\includegraphics[width=1\linewidth]{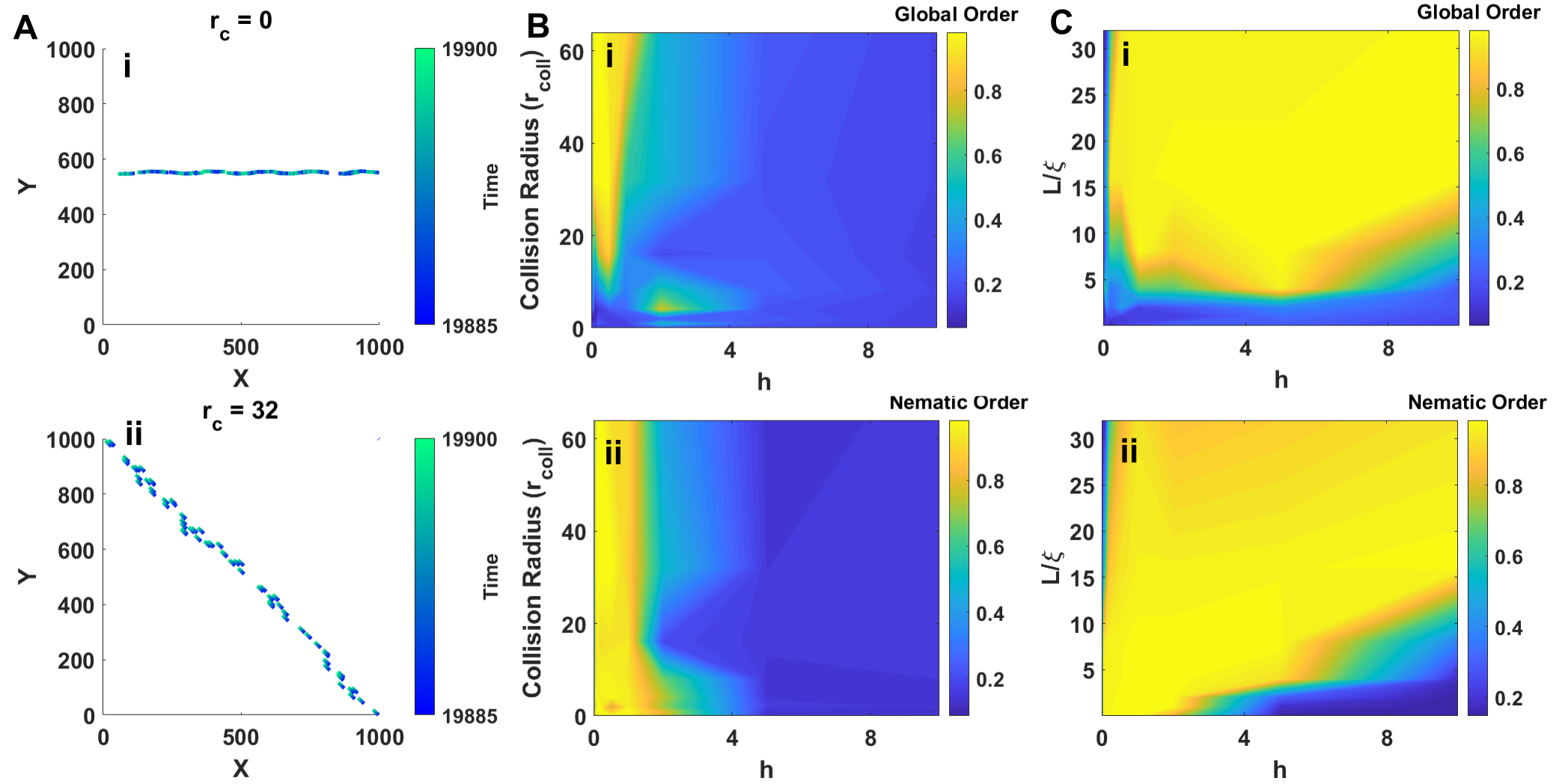}
		\caption{Collective motion in the Harmonic Theory. \textbf{A}: Snapshots of $80$ agents as a function of time and for two different values of  collision radius, $r_{\mathrm{coll}}$, are plotted. For small or zero collision radius, agents form bidirectional columns (\textbf{i}). As the collision radius increases, spontaneous direction selection leads to unidirectional columns with an increased column width to avoid collisions (\textbf{ii}). \textbf{B}: Introducing collision avoidance leads to the formation of unidirectional columns. Global order (angular order parameter in \textbf{i}) and global nematic order (\textbf{ii}) are plotted as a function of collision radius, $r_{\mathrm{coll}}$, and social attraction (the strength of stimuli), $h$. For medium values of $h$, high order is observed. While nematic order is always high in this region, indicating the formation of bidirectional columns, global order is high only for larger values of collision radius, indicating the formation of unidirectional columns due to collision avoidance. \textbf{C}: Introducing distance-dependence of stimuli leads to the formation of unidirectional columns. Global order (\textbf{i}) and global nematic order (\textbf{ii}) as a function of inverse characteristic length of stimuli, normalized by arena size ($L/\xi$), and the stimulus strength are plotted. For small $L/\xi$, the characteristic length of interaction is too high compared to the system size, and interactions are effectively all-to-all (independent of distance). In this regime, the alignment force in the effective Hamiltonian (even harmonics) is symmetric under reflection, and bidirectional columns are observed. When $L/\xi$ increases (i.e., interaction range becomes smaller), the distance dependence of social attraction promotes the formation of unidirectional columns. Parameter values: $N=80$, $v_0^{\mathrm{eff}}=2$, $W=0.9033$, $\sigma=0.1$. Here, up to the first $n_{\mathrm{max}}=64$ harmonics are superimposed. In \textbf{A} and \textbf{B} interactions are distance-independent at long range (attractive with strength $h$), and repulsive with strength $h_{\mathrm{coll}}=-10$ below the collision radius, $r_{\mathrm{coll}}$. In \textbf{A}, $h=0.05$. In \textbf{C}, the amplitude of social stimuli is an exponential, $h\exp(-d/\xi)$, with characteristic length $\xi$, where $d$ is the distance.}
		\label{Fig5}
	\end{figure}
	
	\begin{figure}
		\centering
		\includegraphics[width=1\linewidth]{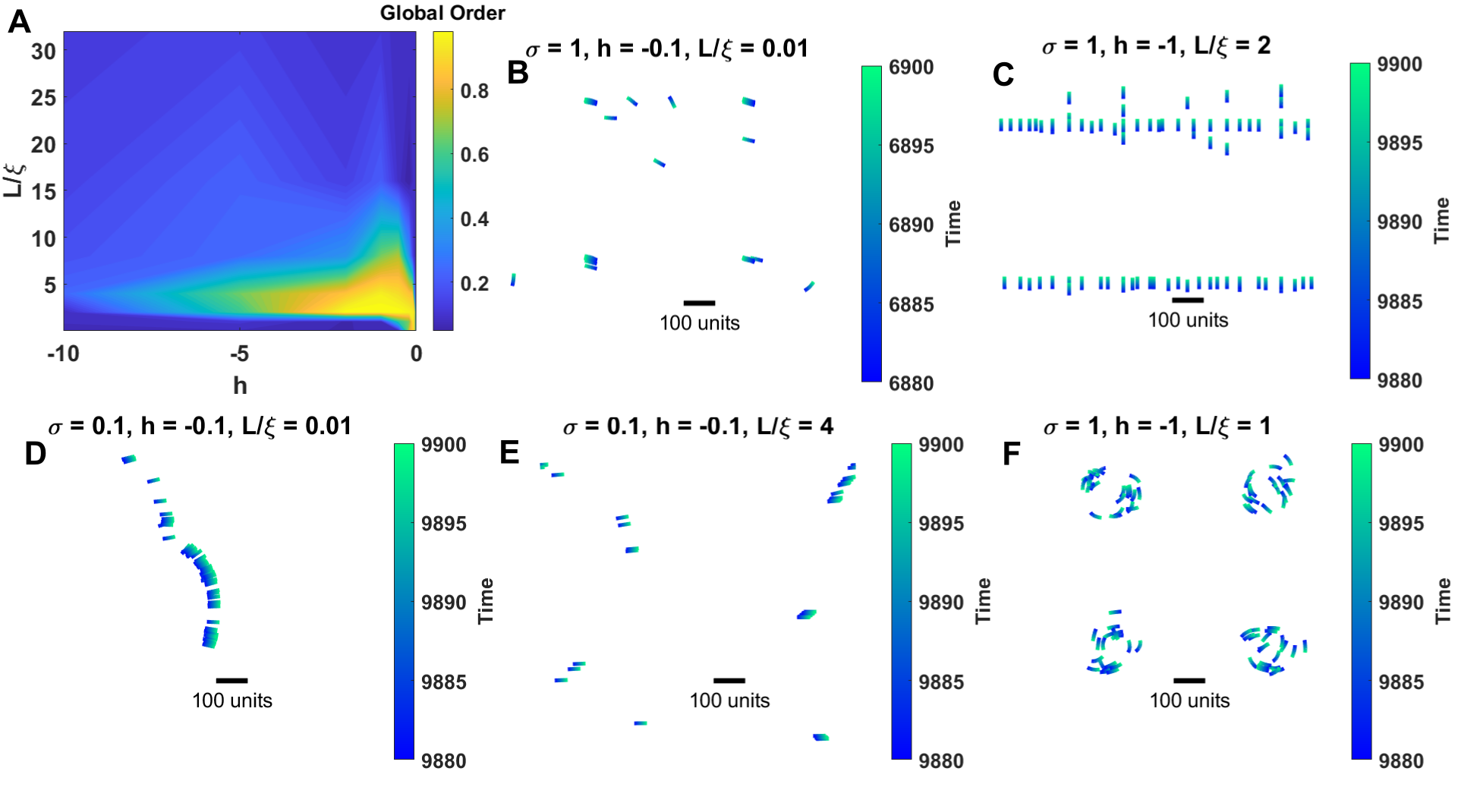}
		\caption{Collective motion resulting from only repulsion in the Harmonic Theory. \textbf{A}: Global order as a function of normalized inverse characteristic length of the stimuli, $L/\xi$, and its strength, $h$, is plotted. Here, the interactions are distance-dependent, according to an exponential, $h\exp(-d/\xi)$, with $h<0$, leading to repulsive stimuli. Global order can reach a high value, indicating that collective motion in the Harmonic Theory can arise from repulsion alone. Collective motion exhibits a wide range of patterns, ranging from subgroup formation, where agents form aligned subgroups performing ballistic motion (\textbf{B}), band formation with group decomposition, where the population forms parallel bands, moving perpendicular to the common bearing of the agents along the band (\textbf{C}), or forming a single band with weak fission-fusion dynamics (\textbf{D}), collective motion with strong fission-fusion dynamics (\textbf{E}), and subgroup formation with rotational milling, where subgroups are located on vertices of a lattice, exhibiting unidirectional or bidirectional rotational milling (\textbf{F}). Parameter values: $N=80$, $v_0^{\mathrm{eff}}=2$, $W=0.9033$, $\sigma=1$ (in \textbf{A}). Here, up to the first $n_{\mathrm{max}}=32$ harmonics are superimposed. Interactions are distance-dependent according to an exponential, $h\exp(-d/\xi$), where $d$ is the distance and $\xi$ is the characteristic length.}
		\label{Fig6}
	\end{figure}
	
	\begin{figure}
		\centering
		\includegraphics[width=1\linewidth]{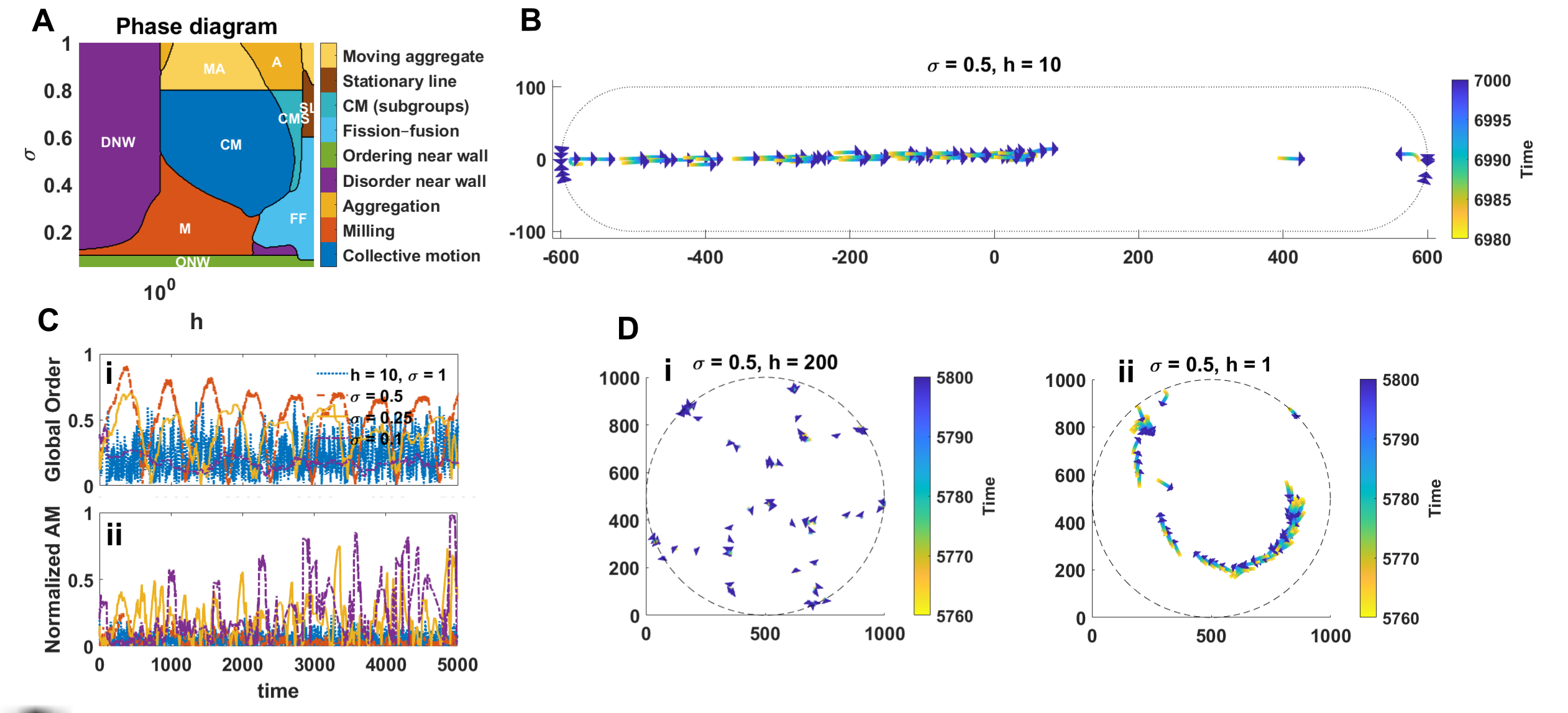}
		\caption{Collective motion in the Harmonic Theory in a bounded space. \textbf{A} to \textbf{C}: The phase diagram of the collective motion in a bounded space (stadium) is plotted (\textbf{A}). For a small sensory kernel, $\sigma$, agents remain nearly stationary close to the walls, leading to a small distance to the walls and a small nearest-neighbor distance. As $\sigma$ increases, for medium $h$, a milling phase, where agents exhibit stronger rotational motion, is observed. However, the rotational motion closely follows the boundary of space, such that angular momentum shows periodic peaks when agents approach curved boundaries where they turn in a coordinated fashion (see \textbf{C} for time series of normalized angular momentum and global order). For yet larger $\sigma$, agents form moving columns going back and forth along the stadium (\textbf{B}), leading to periodic peaks in global order (\textbf{C}). Increasing $h$, agents exhibit aggregates leading to a small nearest-neighbor distance. The aggregate can be stationary or move slowly, giving rise to lower and higher global order, depending on the values of $h$. For too large $h$ and $\sigma$, the aggregate decomposes into subgroups, where the majority of agents exhibit ballistic motion. For too large $h$ and medium $\sigma$, stationary column formation with high ordering and correlated longitudinal and transverse fluctuations along the columns or fission-fusion dynamics with the formation of networked corridors are observed. In this regime, agents naturally form networked corridors in space. \textbf{D}: \textbf{D(i)} and \textbf{D(ii)} show a snapshot of the collective motion pattern, where the population shows stationary networked corridor formation and intermittency between milling and column formation, respectively. Parameter values: $N=80$, $v_0^{\mathrm{eff}}=2$, $W=0.9033$, $h_{\mathrm{coll}}=-1$, and $D_r=0$. Interactions are distance-dependent, attractive according to an exponential with characteristic length $\xi$ ($\xi=125$ in the stadium and $\xi=250$ in the circular arena). Short-range repulsion is included below distance $r_{\mathrm{coll}}$ ($r_{\mathrm{coll}}=64$ in \textbf{A} to \textbf{C} and $r_{\mathrm{coll}}=16$ in \textbf{D}). In \textbf{A} to \textbf{C}, agents move in a rectangular arena with length $1000$ and width $200$, where the two ends are semicircular (half-circle) with radius $100$. In \textbf{D}, agents move in a circular arena with radius $500$. Here, up to the first $n_{\mathrm{max}}=64$ harmonics are superimposed.}
		\label{Fig7}
	\end{figure}

\end{document}